\PassOptionsToPackage{table}{xcolor} 
\documentclass[a4paper,fleqn]{cas-dc}

\usepackage[authoryear,longnamesfirst]{natbib}

\usepackage{amsmath,amssymb,amsfonts}
\usepackage{graphicx}
\usepackage{subcaption} 
\usepackage{booktabs,tabularx}
\usepackage{multirow}
\usepackage{algorithm} 
\usepackage[switch]{lineno}
\usepackage{hhline}
\usepackage{pifont}
\usepackage{url}
\usepackage{soul}
\usepackage{enumitem}
\usepackage{makecell}
\usepackage{etoolbox} 

\definecolor{FoundBlue}{RGB}{220,235,245} 
\definecolor{ReactiveRed}{RGB}{245,220,220} 
\definecolor{ProactiveGreen}{RGB}{220,245,225} 
\definecolor{DiscussPurple}{RGB}{240,225,245} 
\definecolor{FutureYellow}{RGB}{250,245,220} 
\definecolor{arbeige}{RGB}{242, 238, 225} 
\definecolor{arblue}{RGB}{0, 85, 145}

\usepackage{tikz}
\usetikzlibrary{shapes.geometric, arrows, positioning, fit}
\tikzset{
    base/.style={rectangle, rounded corners=4pt, minimum width=6.5cm, minimum height=1.2cm, font=\small, align=center},
    foundation/.style={base, fill=FoundBlue, draw=black, solid, text width=0.59\linewidth},
    reactive/.style={base, fill=ReactiveRed, draw=black, dashed, minimum width=5.5cm, text width=0.46\linewidth},
    proactive/.style={base, fill=ProactiveGreen, draw=black, dotted, minimum width=5.5cm, text width=0.46\linewidth},
    discuss/.style={base, fill=DiscussPurple, draw=black, dashdotted, text width=0.98\linewidth},
    future/.style={base, fill=FutureYellow, draw=black, solid, text width=0.90\linewidth},
    arrow/.style={thick, <->, >=stealth, text width=0.75\linewidth},
    synergy/.style={rectangle, rounded corners=4pt, fill=white, draw=black, font=\small, align=center, minimum width=6cm, minimum height=1cm, text width=0.60\linewidth}
}

\usepackage{caption}
\newcommand{\sectiontitle}[1]{\textbf{#1}\\[-0.8em]\rule{0.96\linewidth}{0.2pt}\\[-0.4em]}

\ExplSyntaxOff
\ExplSyntaxOn
\cs_set:Npn \__first_footerline: {}
\ExplSyntaxOff

\newcommand{\rom}[1]{(\expandafter{\romannumeral #1\relax})}
\makeatletter

\newcommand\citecolor[1]{\@namedef{keycolor#1}{\color{blue}}}

\pretocmd\@bibitem
  {\color{black}\csname keycolor#1\endcsname}
  {}
  {\fail}

\pretocmd\@lbibitem
  {\color{black}\csname keycolor#2\endcsname}
  {}
  {\fail}
\makeatother

\begin{document}

% --- Header and Footer Configuration ---
\let\WriteBookmarks\relax
\def\floatpagepagefraction{1}
\def\textpagefraction{.001}

% Short Title (Header)
\shorttitle{Contingency Planning for Safety-Critical AVs}
% Short Authors (Header)
\shortauthors{Zheng et al.}

% --- Title ---
\title[mode = title]{Contingency Planning for Safety-Critical Autonomous Vehicles: A Review and Perspectives}

% --- Author and Affiliation Information ---

% Affiliation 1: CMU
\affiliation[1]{organization={Robotics Institute, Carnegie Mellon University},
    addressline={5000 Forbes Ave},
    city={Pittsburgh},
    state={PA},
    postcode={15213},
    country={USA}}

% Affiliation 2: HKUST
\affiliation[2]{organization={The Hong Kong University of Science and Technology}, 
    addressline={Clear Water Bay, Kowloon}, 
    city={Hong Kong SAR},
    country={China}}

% Affiliation 3: Delft
\affiliation[3]{organization={Delft University of Technology},
    city={Delft},
    postcode={2628 CD},
    country={Netherlands}}

\author[1,2]{Lei Zheng}
\ead{zack44170625@gmail.com}

\author[3]{Luyao Zhang}
\ead{l.zhang-7@tudelft.nl}

\author[1]{Peiqi Yu}
\ead{peiqiy@andrew.cmu.edu}

\author[1]{Yifan Sun}
\ead{yifansu2@andrew.cmu.edu}

\author[3]{Sergio Grammatico}
\ead{s.grammatico@tudelft.nl}
 
\author[2]{Jun Ma}
\ead{jun.ma@ust.hk}
 
\author[1]{Changliu Liu}
\cormark[1] 
\ead{cliu6@andrew.cmu.edu}

\cortext[1]{Corresponding author}

\begin{abstract}                % Abstract of not more than 250 words. 
Contingency planning is an architectural capability that enables
autonomous vehicles (AVs) { to anticipate and mitigate safety risks arising from safety-relevant logical-mode transitions, such as sensor outages and uncertain interactions}. This paper presents a comprehensive survey of the field, synthesizing fragmented literature into a unified logic-conditioned hybrid control framework. Within this formalism, we categorize approaches into two distinct paradigms:  {Reactive Safety enforces safety constraints or activates fallback
actions after a contingency mode has been observed or reliably inferred,
whereas Proactive Safety preserves recourse across unresolved mode
alternatives through non-anticipative branching.}  In addition, we propose a fine-grained taxonomy that partitions the landscape into external contingencies (environmental and interactive hazards) and internal contingencies (system faults). {Our analysis reveals a fundamental structural divergence: internal faults are predominantly addressed via reactive fail-safe mechanisms, whereas external interaction uncertainties increasingly require proactive branching strategies.} Furthermore, we identify a critical methodological divergence: whereas physical hazards are typically managed with formal guarantees, semantic and out-of-distribution anomalies currently rely heavily on empirical validation. We conclude by identifying the open challenges in bridging the gap between theoretical guarantees and practical validation, advocating for hybrid architectures and standardized benchmarking to transition contingency planning from formulation to certifiable real-world deployment.
\end{abstract}   
\maketitle

% Main text
 
\section{Introduction}
\label{sec:intro}
Autonomous vehicles (AVs) operate in dynamic, unpredictable environments where events like sensor failures or unexpected behaviors of other agents can compromise safety. In such challenging scenarios, human drivers naturally employ contingency reasoning to manage uncertainty. For example, when approaching an occluded crosswalk, a driver may prepare to brake in case a pedestrian unexpectedly emerges. Similarly, to ensure reliable operation under such conditions, AVs must move beyond nominal planning and integrate contingency planning: the capability to anticipate, model, and respond to possible but uncertain deviations from expected operating conditions~\citep{alsterda2021contingency, li2023marc}.  

% We define a contingency as a safety-relevant event that can be formalized as a logic-driven discrete mode switch and is typically low-frequency and high-impact (e.g., "if a pedestrian appears from occlusion").

{We define a contingency as a safety-relevant event or process that can
be represented as a logic-driven transition between discrete logical
modes.} Examples include an actuator malfunction~\citep{alsterda2019contingency}, a vehicle suddenly emerging from an occlusion~\citep{zheng2025oacp}, or an aggressive merge by another driver~\citep{chen2022interactive}. Such events are typically "known unknowns": their structural forms are known at design time, but their occurrence time, likelihood, and trajectory remain uncertain.  {This mode-transition
representation distinguishes contingencies from within-mode continuous
uncertainties, such as small sensor noise.} Contingency planning involves managing modal deviations and logical branches of potential future states, thereby requiring specialized approaches.

The goal of contingency planning is twofold: to ensure safety under such events and to avoid overly conservative behavior that compromises efficiency~\citep{wang2023interactive}. This safety–efficiency trade-off is especially critical for SAE Level 4+ AVs, which operate without human fallback~\citep{gyllenhammar2025road}. Without effective contingency planning, systems may fail when faced with unexpected edge-case events.
 
% --- FIGURE REMAINS THE SAME ---
\begin{figure}
     \centering
     \begin{subfigure}[b]{0.48\linewidth}
         \centering
         \includegraphics[width=\linewidth]{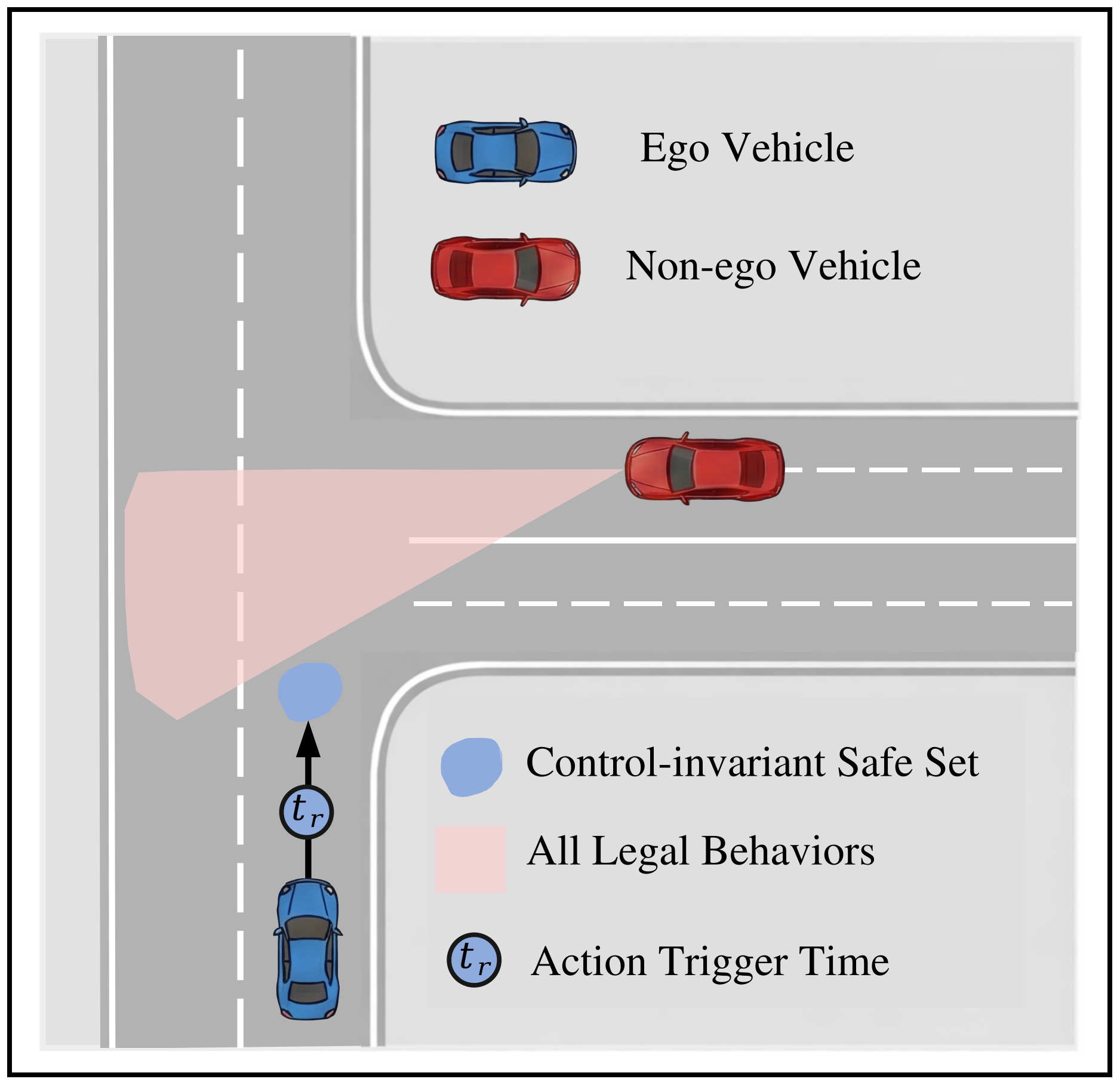}
         \caption{Reactive  Planner}
         \label{fig:a}
     \end{subfigure}
     % \hfill
     \begin{subfigure}[b]{0.48\linewidth}
         \centering
         \includegraphics[width=\linewidth]{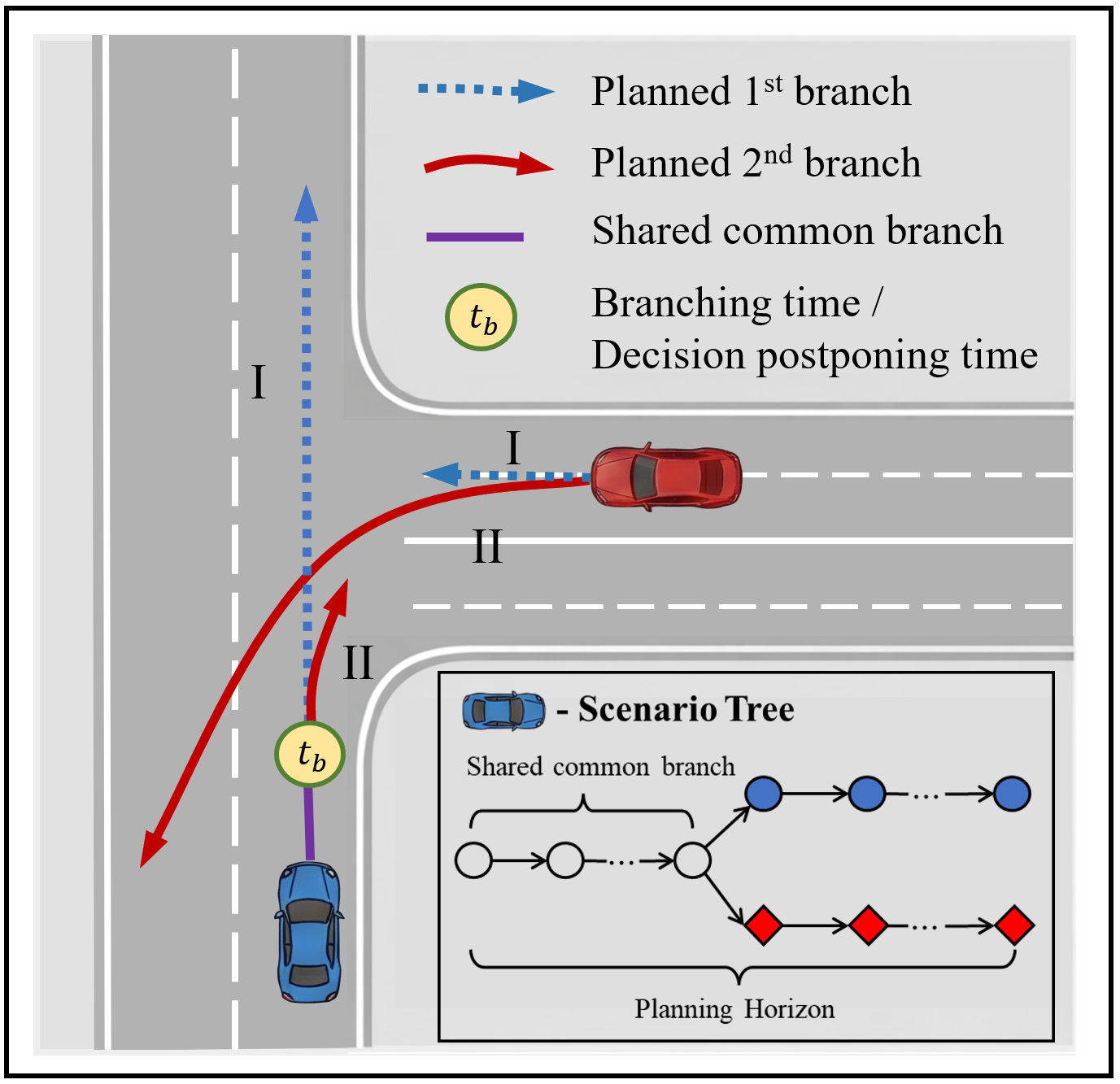}
         \caption{Proactive Planner}
         \label{fig:b}        
     \end{subfigure}
    % \caption{Illustration of Reactive versus Proactive contingency planning for AVs at an urban intersection. (a) The Reactive planner executes a nominal plan until a contingency (the red car's turn) is detected at the action trigger time ($t_r$). It then triggers a safe fallback trajectory that is guaranteed to remain within a control-invariant safe set.(b) The Proactive planner anticipates multiple futures by optimizing a scenario tree from $t=0$. In one scenario, the ego continues if the other vehicle yields (I, dotted path), while in another, the ego stops if the other turns (II, solid path). The common nominal branch is shared, with branching decisions made at the branching time $t_b$, enabling proactive adaptation with lessconservative maneuvers. }
    \caption{{Reactive versus Proactive contingency planning for AVs at an urban intersection. (a) The Reactive planner switches to a fallback trajectory after a turning mode is detected at $t_r$. (b) The Proactive planner preserves mode-conditioned alternatives through a shared trajectory with branching decisions made at the branching time $t_b$.} }
    
%         \caption{
% \textcolor{brown}{Reactive and Proactive contingency planning at an urban intersection.
% The Reactive planner switches to a fallback after the other vehicle's
% turning mode is resolved at $t_r$, whereas the Proactive planner
% preserves mode-conditioned alternatives through a shared trajectory
% until the branching time $t_b$.
% }}
        \label{fig:contingency-fig-urban}
        \vspace{-0pt}
\end{figure}

{This paper organizes contingency planning methods into two primary paradigms},  derived from a unified logic-conditioned control framework based on different assumptions:
\begin{itemize}
\item \textbf{Reactive Safety Methods} (e.g., fail-safe switching), { which condition their intervention on a contingency mode that has been observed or reliably inferred} and trigger predefined policies in response (e.g., "if sensor failure is detected, activate the backup system"). These methods generally rely on real-time monitoring and local policy filtering.

\item \textbf{Proactive Safety Methods} (e.g., branching model predictive control (MPC)), {which anticipate plausible contingency modes while the
associated mode uncertainty remains unresolved.} These methods predict plausible future scenarios by constructing scenario trees and optimizing over them in real time. A shared nominal trunk is maintained, along with multiple branches for proactive adaptation, as shown in Fig.~\ref{fig:b}.
\end{itemize} 

\begin{figure}
     \centering     
    \begin{subfigure}[b]{0.468\linewidth}
         \centering
         \includegraphics[width=\linewidth]{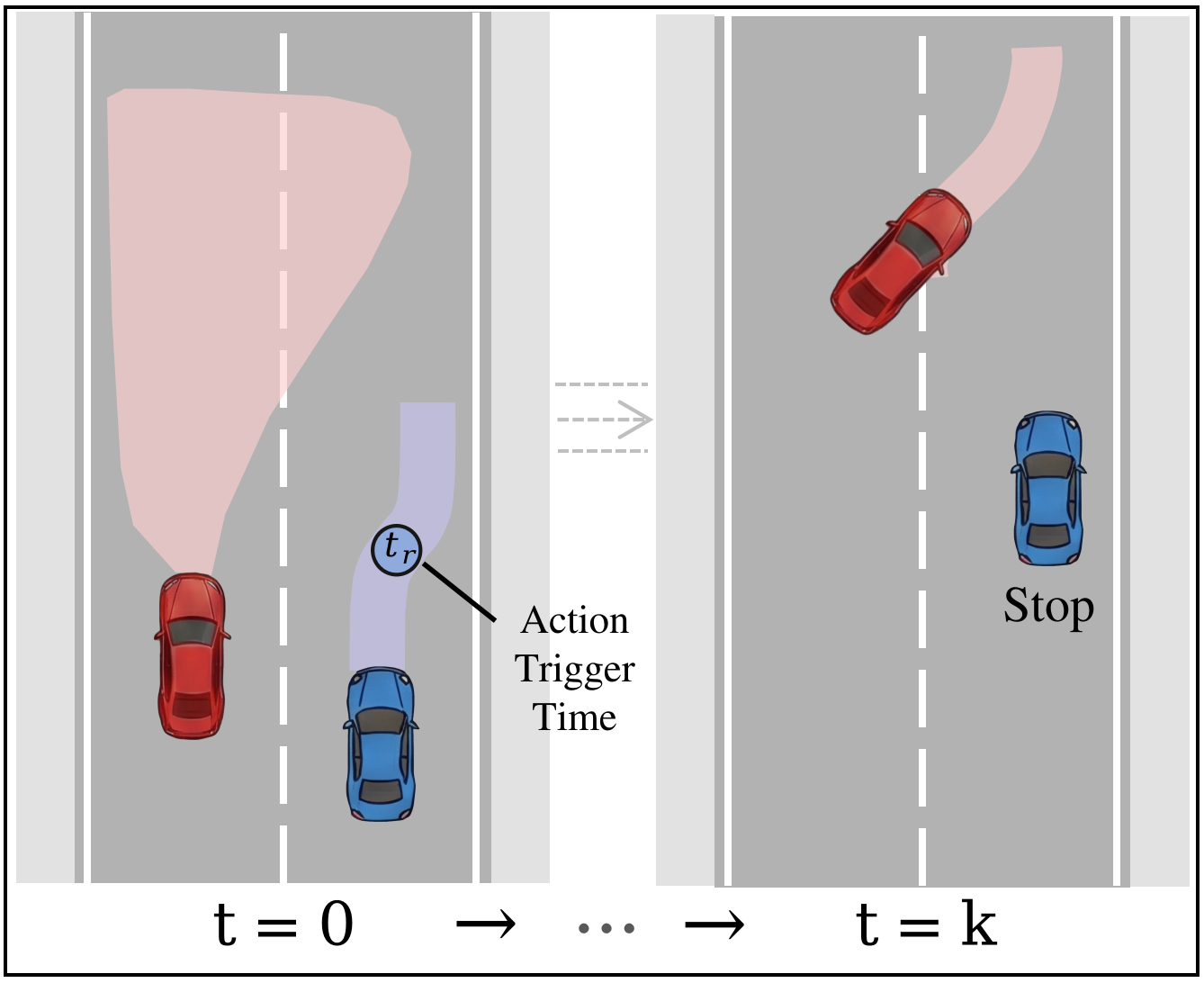}
         \caption{Reactive Planner}
         \label{fig:c}
     \end{subfigure}    
     \begin{subfigure}[b]{0.476 \linewidth}
         \centering
         \includegraphics[width=\linewidth]{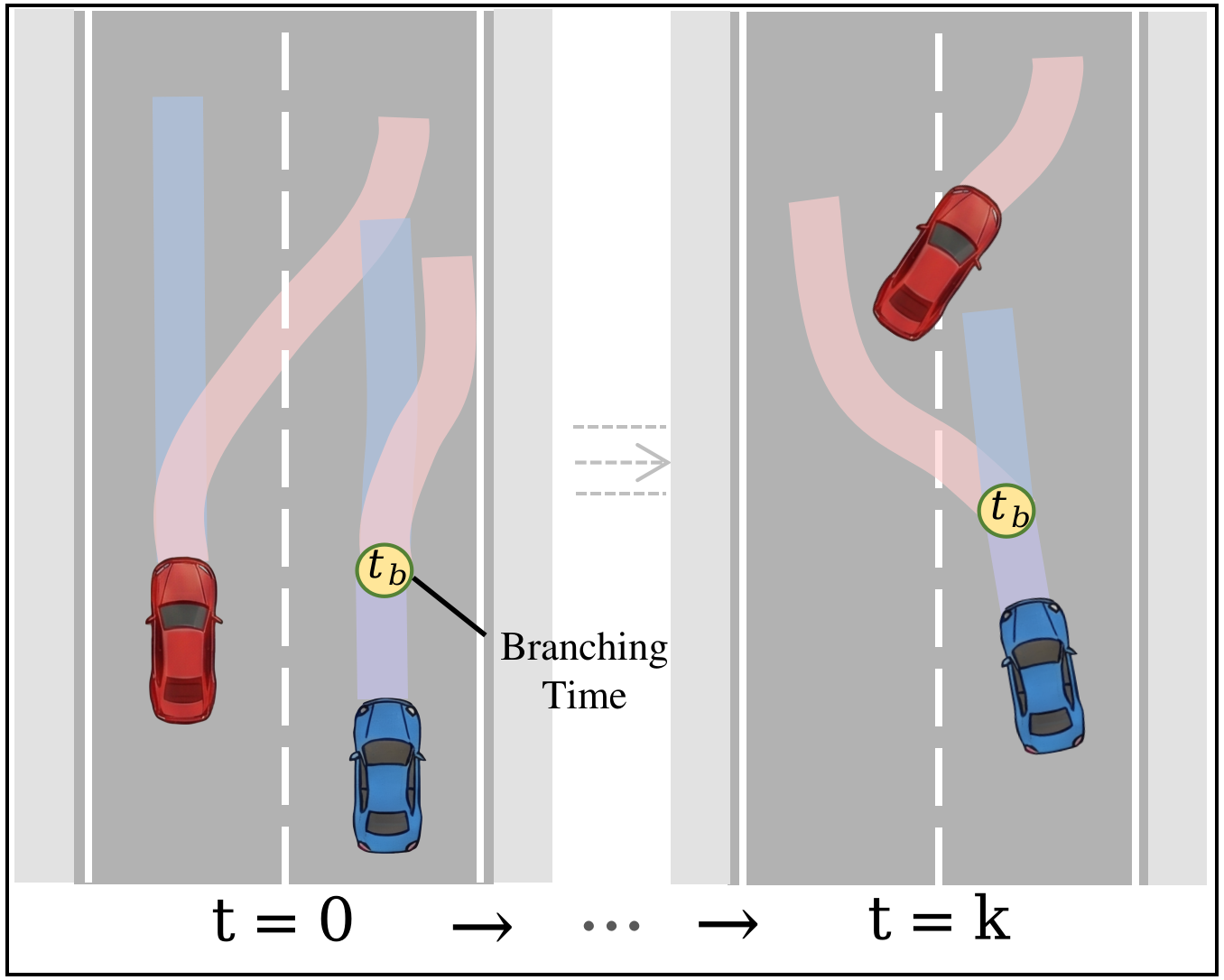}
         \caption{Proactive Planner}
         \label{fig:d}
     \end{subfigure} 
     % \caption{Comparison of Reactive versus Proactive contingency planning during a highway cut-in scenario.(a) The Reactive planner directs the ego vehicle (blue) to execute a conservative stop when the unexpected cut-in is detected at $t_r$.(b) The Proactive planner explicitly reasons over unresolved logical uncertainty from $t=0$. It optimizes a scenario tree of plausible futures (cut-in' and no cut-in') that share a "common trunk". The decision is deferred to the branching time ($t_b$), facilitating a proactive and less conservative maneuver.}
        \caption{
        {Reactive and Proactive contingency planning in a highway cut-in scenario. (a) The Reactive planner directs the ego vehicle (blue) to execute a conservative stop after the cut-in is detected
        at $t_r$. (b) The Proactive planner maintains feasible cut-in and no-cut-in continuations through a shared trajectory until the branching time  $t_b$. 
        }}
\label{fig:contingency-fig-highway} 
        \vspace{-0mm}
\end{figure} 

{
Figures~\ref{fig:contingency-fig-urban} and
\ref{fig:contingency-fig-highway} illustrate this distinction in an urban
intersection and a highway cut-in scenario, respectively. In the Reactive
cases shown in Figs.~\ref{fig:a} and~\ref{fig:c}, the planner follows the
nominal policy until the relevant turning or cut-in mode is observed or
reliably inferred at time $t_r$, after which it applies a
mode-conditioned fallback. In the Proactive cases shown in
Figs.~\ref{fig:b} and~\ref{fig:d}, the planner instead reasons over
multiple unresolved interaction modes by optimizing a scenario tree. The
resulting branches share a common non-anticipative trunk until the
branching time $t_b$, allowing the planner to defer mode-specific
commitment while preserving feasible responses to the modeled outcomes.
In the urban example shown in Fig.~\ref{fig:b}, if the red vehicle yields,
the ego vehicle proceeds through the intersection along branch~I (dotted
path); if the red vehicle turns into the conflict region, the ego vehicle
follows branch~II (solid path) and stops. The highway example applies the
same principle to "cut-in" and "no cut-in" alternatives.
}

{
These examples show that the choice of safety architecture depends not only on the contingency type, but more fundamentally on whether the associated logical mode has been resolved at the time of decision-making. Crucially, these paradigms are not mutually exclusive. Instead, they can form a
complementary cycle}: Proactive planning can help generate {Reactive Safety} certificates, while Reactive monitoring ensures the feasibility of Proactive plans under runtime uncertainty. This synergistic relationship is detailed in Sections~\ref{subsec:connections} and~\ref{subsec:comparative_analysis}.

To advance the development of contingency planning and overcome the current fragmentation in the field, a comprehensive synthesis of existing research is essential. 
{
Existing surveys have provided extensive coverage of automated-driving safety evidence~\citep{gyllenhammar2025road}, fault-tolerance and fallback strategies~\citep{rodriguezarozamena2025fault}, runtime safety filters~\citep{hsu2024safety}, learned safety certificates~\citep{dawson2023safe}, safe learning and reinforcement learning~\citep{brunke2022safe,zhao2023state}, learning-based MPC and MPC-based safety mechanisms~\citep{hewing2020learning}, and Hamilton--Jacobi reachability~\citep{bansal2017hamilton}. These studies are primarily organized around an assurance activity, a methodological family, a learning paradigm, or a specific contingency class. In contrast, this review takes the contingency itself as the organizing principle and represents safety-relevant contingencies through logic-conditioned mode transitions within a unified hybrid control framework. This perspective unifies Reactive methods, which enforce safety after the relevant logical mode has been observed or reliably inferred, with Proactive methods, which reason over unresolved mode alternatives, while jointly covering internal system faults and external environmental and interactive hazards.
This paper presents the first holistic analysis of contingency planning for safety-critical AVs.}    Our key contributions are as follows:
\begin{itemize}
\item  We establish a rigorous mathematical definition of contingency events as discrete, logic-conditioned modal transitions within a stochastic hybrid system. {This representation distinguishes contingencies from standard within-mode continuous uncertainties (e.g., small sensor noise), providing a precise theoretical basis for safety-critical planning.}
% \item We introduced a unified logic-conditioned hybrid control framework that synthesizes disparate methodologies under a single theoretical umbrella. By deriving Reactive safety and Proactive safety as distinct computational approximations of this underlying problem, we elucidate the structural connections and complementary nature of these paradigms.

\item {We adopt a unified logic-conditioned hybrid control framework to synthesize disparate methodologies under a common analytical perspective. By deriving Reactive Safety and Proactive Safety as two tractable solution paradigms of the unified contingency planning problem under distinct logical-mode information assumptions, we elucidate their structural connections and complementary nature.}

\item  We perform a systematic comparative analysis of the trade-offs among safety guarantees, computational tractability, and behavioral conservatism.  {This analysis reveals a recurring structural divergence: while internal system faults are predominantly addressed via Reactive mechanisms, external interaction and semantic uncertainties increasingly motivate Proactive strategies that preserve recourse under
unresolved alternatives. This synthesis provides a conceptual basis for hybrid architectures that integrate the formal safety assurances available
from Reactive filters with recourse-aware Proactive planning
over unresolved alternatives. } 
%  Through this lens, we identify a fundamental "verification gap" in learning-based methods and outline a strategic research roadmap for achieving scalable, certifiable contingency planning in open-world environments.
\end{itemize}

Figure~\ref{fig:roadmap} summarizes the organization and conceptual flow
of this review.
{
Section~\ref{sec:problem_formulation} formalizes contingency planning
within a unified logic-conditioned hybrid control framework.
Sections~\ref{sec:reactive_safety} and~\ref{sec:proactive} review
Reactive and Proactive Safety methods, respectively, distinguished by
whether the safety-relevant logical mode has been resolved.
Section~\ref{sec:discussion} compares their assumptions, guarantees,
trade-offs, and complementary roles, and further discusses closed-loop
mode transitions and dynamic branching.
Finally, Section~\ref{sec:future} summarizes the main findings and
outlines future research directions.
}

\begin{figure*}[t!]
\centering
\begin{tikzpicture}[node distance=1.6cm]
    % 1. Introduction 
    \node (intro) [foundation]
    {\sectiontitle{Section~\ref{sec:intro}. Introduction}
    \begin{itemize}[itemsep=1pt, topsep=0pt]
        \item \textcolor{RoyalBlue}{\textbf{Contribution 1:}} Define contingencies as discrete, logic-conditioned mode transitions
        \item \textcolor{RoyalBlue}{\textbf{Scope:}} Safety-critical contingency planning for AVs
    \end{itemize}};
    
    % 2. Problem Formulation 
    \node (form) [foundation, below of=intro, yshift=-0.5cm] {\sectiontitle{Section~\ref{sec:problem_formulation}. Problem Formulation}
    \begin{itemize}[itemsep=1pt, topsep=0pt]
        \item \textcolor{RoyalBlue}{\textbf{Contribution 2}}: Unified Logic-conditioned hybrid control framework
        \item \textcolor{RoyalBlue}{\textbf{Organizing Criterion}}:
        Is the safety-relevant logical mode resolved?
    \end{itemize}};
    % \textbf{Contribution 2}: Unified hybrid framework\\Two Solution Paradigms};
    
   % 3. Reactive Safety 
    \node (reactive) [reactive, below left of=form, xshift=-3.6cm, yshift=-1.2cm] {\sectiontitle{Section~\ref{sec:reactive_safety}. Reactive Safety Paradigms}
    \begin{itemize}[itemsep=1pt, topsep=0pt]
        \item \textit{{Core Assumption: Logical uncertainty is \textcolor{VioletRed}{\textbf{resolved}}.}}
        \item \textcolor{VioletRed}{{3.1}} Task-Preserving Runtime Filters (CBF, HJ)
        \item \textcolor{VioletRed}{{3.2}} Task-Terminating Fail-Safe Supervision (MRC)
    \end{itemize}};

    % 4. Proactive Safety  
    \node (proactive) [proactive, below right of=form, xshift=3.6cm, yshift=-1.2cm] {\sectiontitle{Section~\ref{sec:proactive}. Proactive Safety Paradigms}
    \begin{itemize}[itemsep=1pt, topsep=0pt]
        \item \textit{Core Assumption: Logical uncertainty is \textcolor{Green}{{unresolved}}.}
        \item \textcolor{Green}{{4.1-4.2}} From Min-Max MPC to Contingency MPC
        \item \textcolor{Green}{{4.3-4.4}} Game-Based  \& Learning-Based Approaches
        \item \textcolor{Green}{{4.5}} Computational Methods
    \end{itemize}
    }; 

    \node (synergy) [synergy, below of=form, yshift=-3.1cm] {\sectiontitle{Connections Between Paradigms}
    \begin{itemize}[itemsep=1pt, topsep=0pt]
        \item {{Proactive→Reactive:}} Non-conservative safety set synthesis
        \item {{Reactive→Proactive:}} Terminal constraint for long-term feasibility
    \end{itemize}
    };
    
    % 5. Discussion 
    \node (discuss) [discuss, below of=synergy, yshift=-1.5cm] {\sectiontitle{Section~\ref{sec:discussion}. Discussion}
     \begin{itemize}[itemsep=1pt, topsep=0pt]
        \item \textcolor{Purple}{\textbf{Contribution 3:}}  Hybrid architecture (Proactive guidance + Reactive filter)
        \item \textcolor{Purple}{{5.1 Comparative Synthesis:}} Safety--conservatism and computation--scalability  \textit{trade-offs} \& Hybrid Integration (Reactive Invariance $\leftrightarrow$ Proactive Recourse)
        \item \textcolor{Purple}{{5.2 Suitability Trends:}}  Internal faults $\rightarrow$ Reactive;
    interactive hazards $\rightarrow$ Proactive;
    semantic/OOD anomalies $\rightarrow$ verification gap 
        \item \textcolor{Purple}{{5.3 Closed-Loop Interaction:}}  Passive prediction versus active interaction-aware contingency planning
        \item \textcolor{Purple}{5.4 Dynamic Branching:}   Heuristic, reachability-based, and deferred-decision strategies
     \end{itemize}
    }; 

    % 6. Future Prospects
    \node (future) [future, below=0.4cm of discuss]
    {\sectiontitle{Section~\ref{sec:future}. Summary \& Future Prospects}
    \begin{itemize}[itemsep=1pt, topsep=0pt]
    
        \item \textcolor{Orange}{\textbf{Core Challenges:}}
        {Scalable contingency modeling;
        verifiable learning-based safety sets; Runtime safety assurance}
    
        \item \textcolor{Orange}{\textbf{Future Directions:}} Technical (physics-data fusion) + Application (social ethics) + Evaluation (standardized benchmarks)
    \end{itemize}
    };

    \draw [thick, ->, >=stealth] (intro) -- (form);
    \draw [thick, ->, >=stealth] (form) -- (reactive);
    \draw [thick, ->, >=stealth] (form) -- (proactive);
    \draw [arrow] (reactive) -- (synergy);
    \draw [arrow] (proactive) -- (synergy);
    \draw [thick, ->, >=stealth] (synergy) -- (discuss);
    \draw [thick, ->, >=stealth] (discuss) -- (future);
\end{tikzpicture}
\caption{ Roadmap of the review on contingency planning for safety-critical AVs. 
{Sections~\ref{sec:intro}--\ref{sec:problem_formulation}
define contingency planning and introduce the unified logic-conditioned
framework. Based on whether the safety-relevant logical mode is resolved,
the literature is organized into Reactive Safety
(Section~\ref{sec:reactive_safety}) and Proactive Safety
(Section~\ref{sec:proactive}). Section~\ref{sec:discussion} synthesizes
their connections, trade-offs, and suitability, while
Section~\ref{sec:future} outlines future research directions.}
}
\label{fig:roadmap} \vspace{-0mm}
\end{figure*}

\section{Problem Formulation}
\label{sec:problem_formulation}
This paper reviews contingency planning for safety-critical AVs. A contingency is a future event or circumstance that is possible but cannot be predicted with certainty~\citep{alsterda2021contingency}. We define a contingency as a "known unknown": a safety-relevant event represented as a stochastic transition in the logical mode $\sigma_k\in\Sigma$ (e.g., the onset of a sensor fault or the resolution of an occlusion).  Notably, while the global safety specification for the system is time-invariant (e.g., no collision), the specific admissible states for the ego vehicle depend on the environment configuration governed by the logical mode. 

\subsection{Preliminaries}
\label{sec:preliminaries}
The ego vehicle is typically governed by the hybrid system:
\begin{subequations}
\label{eq:hybrid_system}
\begin{align}
x_{k+1} &= f_{\sigma_k}(x_k, u_k), \label{eq:cont_dyn}\\
\sigma_{k+1} &= T_k(x_k, \sigma_k, v_k), \label{eq:mode_logic} \\
y_k &= h_{\sigma_k}(x_k), \label{eq:observation}
\end{align}
\end{subequations} where $x_k \in \mathcal{X}$ represents the ego state, $u_k \in \mathcal{U}$ represents the control input,  $y_k \in \mathcal{Y}$ denotes the observation vector (e.g., sensor measurements), and $v_k \in \mathcal{V}$ { represents external factors affecting the
logical-mode transition that are outside the ego vehicle's control, such
as the stochastic onset of a sensor fault or latent characteristics and
decision noise of a surrounding vehicle. $\sigma_k\in\Sigma$ denotes the discrete
safety-relevant logical mode used for contingency reasoning,  such as a
detected sensor fault or a confirmed cut-in. } The logical mode $\sigma_k$ impacts the system in three aspects: (i) First-order dynamics via $f_{\sigma_k}$ (e.g., a transition from nominal friction to a tire blowout, altering the ego vehicle's physical response); (ii) the logical-mode transition map  via $T_k$, which governs environmental changes and safety specifications {(e.g., a transition from an unresolved interaction mode to a conflict-clear mode);}
 and (iii) Observation model via $h_{\sigma_k}$ ({e.g., a sensor fault leading to degraded localization of the ego vehicle}).  
 
Critically, the logical mode $\sigma_k$ often evolves according to transitions that are difficult to predict. {For example, the intent of a surrounding vehicle (e.g., "cutting in" versus "yielding") may change abruptly due to external factors outside the ego
vehicle's control and may not be reliably forecastable from available
measurements. 
Consequently, a fundamental challenge in contingency planning is
managing the mismatch between the actual evolution of the logical mode
and the mode transitions assumed by the controller.}

{In this paper,  $\bar{\sigma} \in \Sigma$ denotes a
fixed mode value used in mode-conditioned definitions. For each
$\bar{\sigma} \in \Sigma$, we define the mode-dependent safe set and
failure set as
\[
\mathcal{X}_{\mathrm{safe},\bar{\sigma}}
\subseteq \mathcal{X},
\qquad
\mathcal{F}_{\bar{\sigma}}
:=
\mathcal{X}
\setminus
\mathcal{X}_{\mathrm{safe},\bar{\sigma}},
\]
respectively. A mode-dependent control-invariant subset, when used, is
denoted by $\mathcal{X}_{\mathrm{inv},\bar{\sigma}}
\subseteq
\mathcal{X}_{\mathrm{safe},\bar{\sigma}}$, 
and is formally defined in Section~\ref{subsec:reactive_paradigm}.
A scenario over a horizon $N$ is a finite mode sequence
$(\sigma_0,\ldots,\sigma_N)$.}

\begin{figure}
     \centering     \includegraphics[width=0.975\linewidth]{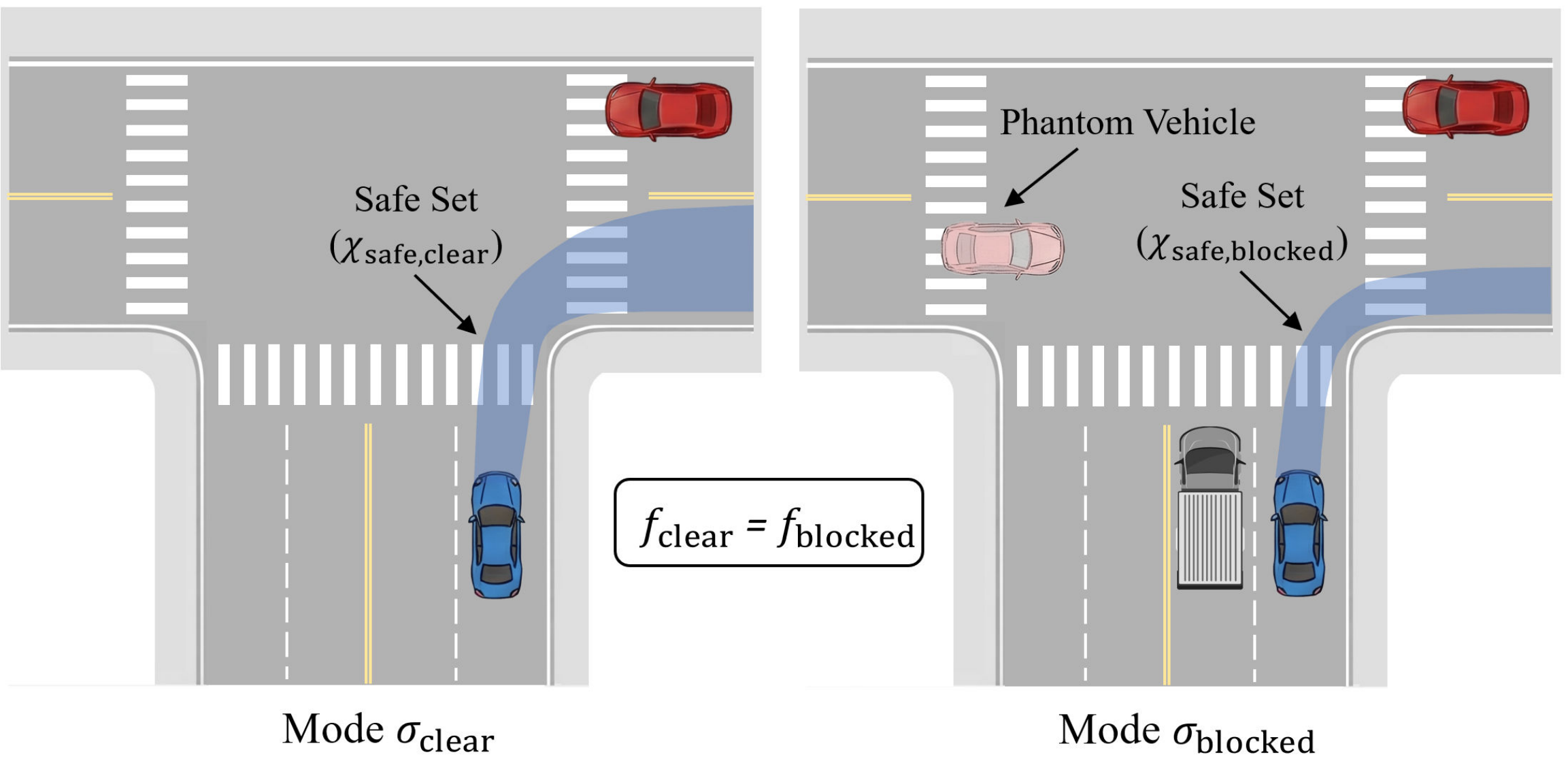}
        % \caption{Visual illustration of how contingency modes $\sigma$ impact the ego safety under occlusion. An environmental obstruction creates specification uncertainty. The dynamics $f$ remain nominal (reachable set is unchanged), but the Safe Set shrinks ($\mathcal{X}_{\text{safe},\text{blocked}} \subset \mathcal{X}_{\text{safe},\text{clear}}$) to account for potential hidden hazards in the occluded region.}
        \caption{{Visual illustration of how the fixed contingency mode
        $\bar{\sigma}$ affects ego-vehicle safety under occlusion. An
        environmental obstruction creates uncertainty in the safety
        specification. The continuous dynamics remain unchanged,
        $f_{\text{clear}} = f_{\text{blocked}}$, whereas the admissible safe
        set contracts according to
        $\mathcal{X}_{\mathrm{safe},\text{blocked}}
        \subset
        \mathcal{X}_{\mathrm{safe},\text{clear}}$
        to account for potential hidden hazards in the occluded region.}}
\label{fig:contingency-mode-occ} 
        \vspace{-0pt}
\end{figure}   

\begin{figure}
     \centering     \includegraphics[width=0.975\linewidth]{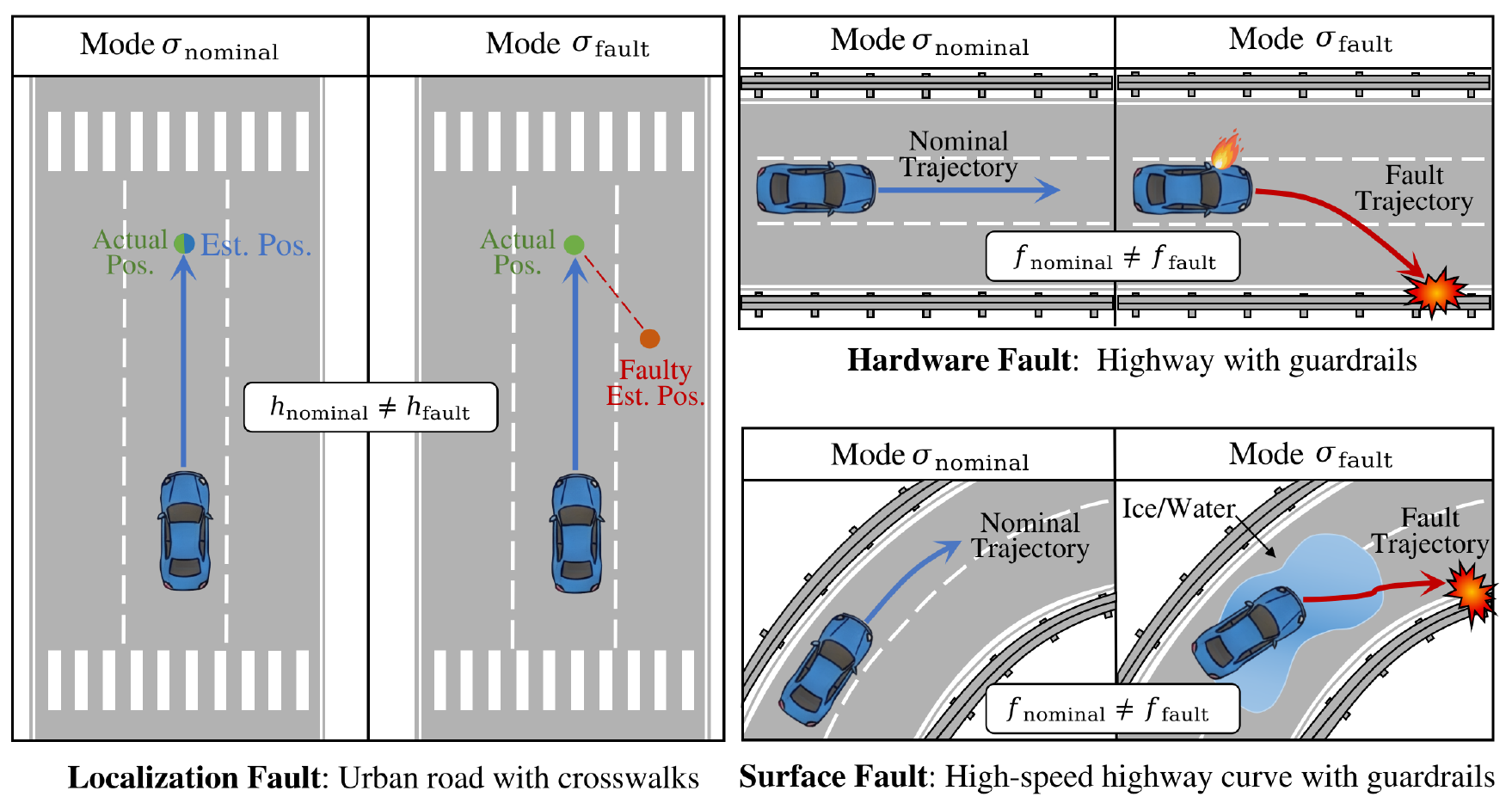}
        % \caption{Visual illustration of how contingency modes $\sigma$ impact the ego vehicle under different contingency events. }
        \caption{{Visual illustration of how different fixed contingency modes
        $\bar{\sigma} \in \Sigma$ affect the ego vehicle through changes in
        the dynamics, safety specification, or observation model.}}
\label{fig:contingency-mode-others} 
        \vspace{-0pt}
\end{figure}

{The safe set
$\mathcal{X}_{\mathrm{safe},\bar{\sigma}}$
represents the static admissibility constraints, such as
non-collision, associated with the fixed mode
$\bar{\sigma}$. For example, an occlusion scenario can be represented
by
\(
\bar{\sigma}
\in
\{\text{clear},\text{blocked}\},
\)
with identical continuous dynamics,
$f_{\text{clear}} = f_{\text{blocked}}$, but different safety
specifications. The blocked mode,
$\bar{\sigma}=\text{blocked}$, introduces a latent constraint, such as
a potential hidden actor, and therefore defines the tightened safe set
\(
\mathcal{X}_{\mathrm{safe},\text{blocked}}
\subset
\mathcal{X}_{\mathrm{safe},\text{clear}},
\)
as illustrated in Fig.~\ref{fig:contingency-mode-occ}. Dynamics faults,
such as a loss of road friction, can similarly be represented by
\(
\bar{\sigma}
\in
\{\text{nominal},\text{fault}\},
\)
where $f_{\text{nominal}}$ represents normal operation and
$f_{\text{fault}}$ represents degraded actuation. Such a mode switch
alters the system's reachability rather than necessarily changing the
safe-set boundary.}
More illustrations are shown in Fig.~\ref{fig:contingency-mode-others}.
% Finally, contingencies can directly impact the observation model $h_{\sigma}$; a sensor failure mode implies that the observation $y_k$ no longer accurately reflects the state $x_k$ (or environment configuration), distinct from nominal operation 

\subsection{The Contingency Planning Problem}
\label{subsec:unified_problem}
The core contingency planning problem is to find a policy that optimizes performance while ensuring that safety requirements are satisfied under the hybrid evolution~\eqref{eq:hybrid_system}. To express safety in a way that explicitly couples logical conditions with continuous constraints, we use a temporal-logic specification. Let \(\phi_{\text{safe}}\) be the safety formula
% \begin{equation}
%     \phi_{\text{safe}} \triangleq \bigwedge_{\sigma \in \Sigma} \mathbf{G}_{[0, \infty)} \left( \sigma_k = \sigma \;\Rightarrow\; x_k \in \mathcal{X}_{\mathrm{safe},\sigma} \right),
%     \label{eq:stl_spec}
% \end{equation}
% where \(\mathbf{G}_{[0, \infty)}\) denotes the temporal "Globally" operator over an infinite horizon~\citep{baier2008principles}. This condition states that whenever the system is in mode \(\sigma\), its state must lie in the corresponding safe set.
{\begin{equation}
    \phi_{\mathrm{safe}}
    \triangleq
    \bigwedge_{\bar{\sigma} \in \Sigma}
    \mathbf{G}_{[0,\infty)}
    \left(
    \sigma_k = \bar{\sigma}
    \;\Rightarrow\;
    x_k
    \in
    \mathcal{X}_{\mathrm{safe},\bar{\sigma}}
    \right),
    \label{eq:stl_spec}
\end{equation}
where $\mathbf{G}_{[0,\infty)}$ denotes the temporal "Globally"
operator over an infinite horizon. This condition states that whenever
the current mode $\sigma_k$ equals a fixed mode value
$\bar{\sigma}$, the state must lie in the corresponding
mode-dependent safe set
$\mathcal{X}_{\mathrm{safe},\bar{\sigma}}$.}

Let \(\mathcal{I}_k\) denote the information set available at time \(k\) (e.g., history of observations). A causal policy \(\pi\) maps \(\mathcal{I}_k\) to control inputs \(u_k = \pi(\mathcal{I}_k)\). The contingency planning problem is formulated as
\begin{subequations}
\label{eq:unified_problem_logical}
\begin{align}
  \min_{\pi}\;\; & J(\pi) \label{eq:unified_obj}\\
  \text{s.t.}\quad
  & (x_{k+1},\sigma_{k+1})\, \text{ follow \eqref{eq:hybrid_system} with } u_k = \pi(\mathcal{I}_k), \label{eq:unified_dyn}\\
  & x_k \in \mathcal{X},\;\; u_k \in \mathcal{U}, \qquad \forall k, \label{eq:unified_cons}\\
  & \mathbb{P}\!\left[(x_0,\sigma_0), (x_1,\sigma_1), \dots \models \phi_{\text{safe}}\right] \ge 1 - \epsilon, \label{eq:unified_safety}
\end{align}
\end{subequations}
where the probability is taken over the realization of \(v_k\), and \(\epsilon \in [0,1]\) is a user-specified risk tolerance. \(\models\) denotes the satisfaction relation between a trace and a formula. Condition~\eqref{eq:unified_safety} enforces safety over an infinite horizon, while \eqref{eq:stl_spec} makes the logical structure explicit.

{
The probabilistic formulation should not be interpreted as using the
logical mode to encode the numerical likelihood of a behavior. For
interaction contingencies, the logical mode represents whether the behavior is committed, such as
"unknown", "committed cut-in", or "committed non-cut-in". While
the behavior remains unknown, Proactive reasoning
(Section~\ref{subsec:proactive_paradigm}) is required regardless of
whether the estimated probability of a cut-in is relatively high or
low, because both committed outcomes remain possible.}

{The framework is nevertheless compatible with probabilistic behavior
modeling: predicted probabilities may guide scenario construction,
branch weighting, and risk-sensitive objectives. In practice, an
application-dependent confidence threshold may provide an operational
rule for inferring commitment. For example, the interaction may be
treated as a committed cut-in or committed non-cut-in when the posterior
probability of the corresponding outcome exceeds an illustrative
threshold of $0.95$; otherwise, it remains unresolved. Any subsequent
Reactive treatment
(Section~\ref{subsec:reactive_paradigm}) based on this inference remains
conditional on the calibration and reliability of the upstream
inference process.
}

{Solving~\eqref{eq:unified_problem_logical} exactly is intractable due to the stochasticity of \(v_k\)  and the combinatorial complexity of mode sequences.} Therefore, the two predominant paradigms are distinguished by how the logical uncertainty is addressed in~\eqref{eq:mode_logic} when enforcing~\eqref{eq:unified_safety}.

\subsection{Solution Paradigm I: Reactive Safety}
\label{subsec:reactive_paradigm}
% This paradigm is derived from the problem \eqref{eq:unified_problem_logical} under {a critical assumption: the logical uncertainty is resolved. Specifically, it assumes the discrete mode $\sigma_k$ is known or reliably estimated (e.g., a detected fault or observed occlusion).} 

% Under this assumption, the contingency problem collapses into a local, deterministic enforcement problem. The algebraic structure of the supervisor is given by \eqref{eq:reactive_filter},  comprising a nominal controller $\pi^{\mathrm{nom}}$ and a supervisory policy $\pi^{\ell}$ that filters the nominal command:  
% \begin{equation}
% u_k = \pi^{\ell}\big(x_k,\pi^{\mathrm{nom}}(\mathcal{I}_k),\sigma_k\big).
% \label{eq:reactive_filter}
% \end{equation}

% For each mode $\sigma \in \Sigma$, the supervisor is designed to ensure that the state remains within a mode-dependent control-invariant set $\mathcal{X}_{\mathrm{inv},\sigma} \subseteq \mathcal{X}_{\mathrm{safe},\sigma}$ satisfying:
% \begin{equation}
% \forall\, x \in \mathcal{X}_{\mathrm{inv},\sigma}\;\; \exists\, u \in \mathcal{U} \;\text{s.t.}\;  
%  f_{\sigma}(x,u) \in \mathcal{X}_{\mathrm{inv},\sigma}.
% \label{eq:invariance}
% \end{equation} 

{This paradigm is derived from
\eqref{eq:unified_problem_logical} under the critical assumption that
the logical uncertainty is resolved. Specifically, it assumes the discrete mode $\sigma_k$ is known or reliably inferred (e.g., a detected fault). }

{Under this assumption, the contingency problem reduces to a local, mode-conditioned enforcement problem}. The algebraic structure of the supervisor is given by \eqref{eq:reactive_filter},  comprising a nominal controller $\pi^{\mathrm{nom}}$ and a supervisory policy $\pi^{\ell}$ that filters the nominal command:  
{\begin{equation}
u_k =
\pi^{\ell}
\big(
x_k,
\pi^{\mathrm{nom}}(\mathcal I_k),
\bar{\sigma}
\big).
\label{eq:reactive_filter}
\end{equation}}

{For each fixed mode value $\bar{\sigma} \in \Sigma$, the supervisor is
designed to keep the state within a mode-dependent control-invariant
set $\mathcal{X}_{\mathrm{inv},\bar{\sigma}}
\subseteq
\mathcal{X}_{\mathrm{safe},\bar{\sigma}}$, 
satisfying
\begin{equation}
\forall x
\in
\mathcal{X}_{\mathrm{inv},\bar{\sigma}},
\quad
\exists u \in \mathcal{U}
\quad
\text{s.t.}
\quad
f_{\bar{\sigma}}(x,u)
\in
\mathcal{X}_{\mathrm{inv},\bar{\sigma}}.
\label{eq:invariance}
\end{equation}}
\subsection{Solution Paradigm II: Proactive Safety}\label{subsec:proactive_paradigm} 
{
Proactive methods tackle \eqref{eq:unified_problem_logical} when the
relevant logical uncertainty remains unresolved. In this case, several
mode alternatives remain compatible with the available information, and
the scenario tree represents their corresponding mode-conditioned future
continuations.
}
Instead of conditioning on a single resolved mode, this paradigm
computes a tractable finite-horizon approximation of
\eqref{eq:unified_problem_logical} {by constructing a scenario
tree $\mathcal{T}$ and optimizing a non-anticipative policy over its
branches. Let $\mathcal{B}(\mathcal{T})
=
\{1,\ldots,S\} $ denote the set of root-to-leaf branches of $\mathcal{T}$. Each branch
$b \in \mathcal{B}(\mathcal{T})$ is associated with the discrete mode sequence
\begin{equation}
\boldsymbol{\sigma}_{0:N}^{b}
=
(\sigma_0^b,\ldots,\sigma_N^b),
\qquad
\sigma_k^b \in \Sigma, \notag
\end{equation}
where $\sigma_k^b$ denotes the mode at time $k$ along branch $b$.} {
Each branch represents a mode sequence that remains compatible with the
information available at the planning time and is admissible under the
logical transition map $T_k$.
The corresponding state and control variables are denoted by
$x_k^b$ and $u_k^b$, respectively. 
The Proactive problem is then formulated as a Constrained Branching Decision Process (CB-DP)}:
{\begin{subequations}
\label{eq:rcbdp_problem}
\begin{align}
\hspace{-8mm}
\min_{\{u_k^b\}}
\quad
&
\sum_{b \in \mathcal{B}(\mathcal{T})}
w_b\left(
\sum_{k=0}^{N-1}
c(x_k^b,u_k^b)
+
V_f(x_N^b)
\right)
\label{eq:rcbdp_obj}\\
\hspace{-8mm}
\text{s.t.}
\quad
&
x_0^b=x_0,
\quad
\forall b \in \mathcal{B}(\mathcal{T}),
\label{eq:rcbdp_init}\\ 
&
x_{k+1}^b
=
f_{\sigma_k^b}(x_k^b,u_k^b),
\quad
\forall k,b,
\label{eq:rcbdp_dyn}\\
&
x_k^b
\in
\mathcal{X}_{\mathrm{safe},\sigma_k^b},
\quad
u_k^b \in \mathcal{U},
\quad
\forall k,b,
\label{eq:rcbdp_state_constraint}\\
&
u_k^b=u_k^{b'},
\quad
\forall k<\tau(b,b'),  \quad \text{(non-anticipativity)}, 
\label{eq:rcbdp_nonant}\\
&\forall  b,b' \in \mathcal{B}(\mathcal{T}). \nonumber
\end{align}
\end{subequations}} 
{Here, $w_b\geq 0$ denotes the weight assigned to branch $b$. 
For example, it may be derived from a probabilistic behavior model. These
weights affect the optimization objective but do not determine whether
the logical mode is resolved.  \(c(\cdot,\cdot)\)  and \(V_f(\cdot)\) denote the stage cost and the terminal cost, respectively.} 
{The quantity $\tau(b,b')$ denotes the first time at which branches
$b$ and $b'$ become distinguishable from the available information
$\mathcal{I}_k$. The hard safety constraint
\eqref{eq:rcbdp_state_constraint} enforces
finite-horizon constraint satisfaction along every modeled branch. Such branch-wise constraints provide sample-path guarantees
for the scenarios represented by the tree and approximate
\eqref{eq:unified_safety} according to the coverage of
$\mathcal{B}(\mathcal{T})$. The non-anticipativity constraint \eqref{eq:rcbdp_nonant} couples
branches that share the same information history. Specifically, it
enforces identical decisions whenever
$k<\tau(b,b')$, thereby creating the "common nominal trunk", as shown in Figures~\ref{fig:b} and~\ref{fig:d}.}

{In a receding-horizon implementation, only the first control action on the shared trunk segment is executed before the tree is updated using newly available observations. This facilitates decision consistency while reasoning over multiple unresolved futures before the branches
become distinguishable.
Once the relevant mode is resolved and the corresponding branches become
distinguishable from the available information, their mode-conditioned
controls may diverge.
}

\subsection{{Connections Between Paradigms}}
\label{subsec:connections}
% As approximations of the contingency problem \eqref{eq:unified_problem_logical}, the Reactive \eqref{subsec:reactive_paradigm} and Proactive \eqref{subsec:proactive_paradigm} paradigms are not competing. Instead, each provides a critical component that addresses a fundamental weakness in the other.  

% The effectiveness of the Reactive supervisor \eqref{eq:reactive_filter} hinges on the quality of its pre-computed "safety critic" (e.g., the invariant set $\mathcal{X}_{\mathrm{inv},\sigma}$ in \eqref{eq:invariance}). The synthesis of a high-quality, non-conservative set $\mathcal{X}_{\mathrm{inv},\sigma}$ is a complex, long-horizon optimal control problem that must account for all future modes. 
As approximations of the contingency problem
\eqref{eq:unified_problem_logical}, the Reactive and Proactive paradigms are complementary rather than competing. Each addresses a limitation of the other.

The effectiveness of the Reactive supervisor
\eqref{eq:reactive_filter} hinges on the quality of its precomputed
"safety critic", {such as the mode-dependent invariant set
$\mathcal{X}_{\mathrm{inv},\bar{\sigma}}$ defined in
\eqref{eq:invariance}. Synthesizing a high-quality,
non-conservative set
$\mathcal{X}_{\mathrm{inv},\bar{\sigma}}$
is a complex long-horizon optimal-control problem that must account
for the dynamics $f_{\bar{\sigma}}$, admissible disturbances, input
constraints, and the corresponding safety specification.}
This offline synthesis is itself a look-ahead/proactive computation. Tools such as Hamilton-Jacobi (HJ) reachability~\citep{borquez2024safety} or an offline version of the scenario-based optimization \eqref{eq:rcbdp_problem}  can compute the invariant sets used by computationally efficient online Reactive filters.

Conversely, the Proactive method \eqref{eq:rcbdp_problem}, {as a finite-horizon approximation, does not by itself guarantee safety and recursive feasibility beyond the horizon $N$.} The Reactive paradigm provides the formal solution. 
% By using the control-invariant set $\mathcal{X}_{\mathrm{inv},\sigma}$ from the reactive formulation \eqref{eq:invariance} as a terminal constraint in the proactive optimization, we guarantee recursive feasibility. The state constraints \eqref{eq:rcbdp_state_constraint} are strengthened with:
% \begin{equation}
% x_{N,s} \in \mathcal{X}_{\mathrm{inv},\sigma_{N,s}}, \quad \forall s \in \mathcal{S}_{\mathrm{tree}}.
% \label{eq:bridge_terminal}
% \end{equation}
{By imposing the mode-dependent control-invariant set from
\eqref{eq:invariance} as a terminal constraint on each branch of the
Proactive optimization, recursive feasibility can be established
under the corresponding assumptions. Specifically, the state
constraints \eqref{eq:rcbdp_state_constraint} are strengthened with
\begin{equation}
x_N^b
\in
\mathcal{X}_{\mathrm{inv},\sigma_N^b},
\qquad
\forall b \in \mathcal{B}(\mathcal{T}).
\label{eq:bridge_terminal}
\end{equation}} 
This constraint \eqref{eq:bridge_terminal} ensures that every branch of the Proactive plan terminates in a "safe anchor" state, {from which the Reactive supervisor \eqref{eq:reactive_filter} preserves safety beyond the planning horizon.}

 % To characterize the fundamental limit of this safety guarantee, we define the backward reachable set of $\mathcal{F}_{\sigma}$, denoted $BRS(\mathcal{F}_{\sigma})$, as the \textit{inevitable failure set}: the set of states from which entry into the failure set $\mathcal{F}_{\sigma}$ is unavoidable under any admissible control~\citep{mitchell2005time}. Additionally, the complement of this inevitable failure set constitutes the maximal control-invariant set \citep{blanchini1999set}. Notably, Reactive safety will fail if the contingency happens in the region inside the inevitable failure set $ BRS(\mathcal{F}_{\sigma})$.
 {For a fixed mode $\bar{\sigma}$, let $\operatorname{BRS}_{\bar{\sigma}}
\big(
\mathcal{F}_{\bar{\sigma}}
\big) $
denote the backward reachable set of the failure set
$\mathcal{F}_{\bar{\sigma}}$ under the mode-conditioned dynamics
$f_{\bar{\sigma}}$. We refer to this set as the
\textit{inevitable failure set}: it contains the states from which
entry into $\mathcal{F}_{\bar{\sigma}}$ is unavoidable under every
admissible control. Under the standard reachability assumptions, its
complement corresponds to the maximal control-invariant safe set
$\mathcal{X}_{\mathrm{inv},\bar{\sigma}}^{\max}$.  Notably,  the Reactive planner cannot prevent failure if a contingency is
detected only after the state has entered
$\operatorname{BRS}_{\bar{\sigma}}
(\mathcal{F}_{\bar{\sigma}})$.}
 Proactive planning can provide guidance by anticipating mode transitions and steering trajectories to avoid the inevitable failure set along the horizon, while \eqref{eq:bridge_terminal} anchors each branch in a control-invariant set.

 \section{Reactive Safety Paradigms}\label{sec:reactive_safety}
 % \hl{TODO:}  core idea, mathematical principles and solution strategy, concrete examples in a variety of AV contingency scenarios, corresponding methods to these scenarios, open and unsolved problems.
The {Reactive Safety} architecture employs hierarchical policies to ensure AVs maintain or transition to a safe state following a contingency. This paradigm operates under the assumption that the logical uncertainty has been resolved,  {meaning that the relevant
logical-mode trajectory over the planning horizon is known or reliably
inferred. The resolved trajectory need not be constant. For example, if
it is known that the mode will remain "clear" for two seconds and
then switch to "occupied", the problem remains Reactive because no
unresolved mode alternative must be considered. In practice, Reactive formulations often adopt the common special case that the currently identified mode persists over the local enforcement horizon.}

{For this common fixed-mode case, suppose that the mode
observed at the intervention time $t$ is
$\sigma_t=\bar{\sigma}$ and that this resolved mode is assumed to
persist over the local enforcement horizon. The fundamental principle
is recursive feasibility: an admissible control sequence must exist
that keeps all future states within
$\mathcal{X}_{\mathrm{safe},\bar{\sigma}}$. We formalize this property
through the mode-dependent control-invariant set
$\mathcal{X}_{\mathrm{inv},\bar{\sigma}}$ satisfying
\eqref{eq:invariance}. If $x_t
\in
\mathcal{X}_{\mathrm{inv},\bar{\sigma}}$, 
then an admissible policy guarantees
\[
x_k
\in
\mathcal{X}_{\mathrm{inv},\bar{\sigma}}
\subseteq
\mathcal{X}_{\mathrm{safe},\bar{\sigma}},
\qquad
\forall k\geq t.
\]
Within this framework, a nominal policy $\pi^{\mathrm{nom}}$
prioritizes task performance, while a safety supervisor $\pi^{\ell}$
monitors the state and intervenes according to the mode-conditioned
dynamics. The supervised input
\(
u_k
=
\pi^{\ell}
\big(
x_k,
\pi^{\mathrm{nom}}(\mathcal{I}_k),
\bar{\sigma}
\big)
\)
is synthesized to ensure
\(
x_{k+1}
\in
\mathcal{X}_{\mathrm{inv},\bar{\sigma}},
\)
thereby preserving invariance.}
We categorize concrete supervisor realizations into two classes: task-preserving runtime filters (Section~\ref{subsec:runtime_filter}) and task-terminating fail-safe supervision (Section~\ref{subsec:fail_safe}).

\subsection{Task-Preserving Runtime Safety Filters} 
\label{subsec:runtime_filter}
\subsubsection{Core Idea: The Unified Value Function}
Task-preserving filters instantiate the supervisor $\pi^{\ell}$ to monitor a performance-oriented policy $\pi^{\mathrm{nom}}$, intervening to modify commands only when necessary. 
% For safety-critical AVs, a natural unifying viewpoint is to encode safety through a scalar \textit{safety value function}  {$V_{\sigma}(x): \mathcal{X} \to \mathbb{R}$. For a fixed mode $\sigma$,} the zero-superlevel set of this function, $\mathcal{S}_{{\sigma}} := \{x \in \mathcal{X} \mid V_{{\sigma}}(x) \ge 0\} \subseteq \mathcal{X}_{\mathrm{safe},{\sigma}}$, defines the region of safe operation. The core mechanism of the filter is to enforce the forward invariance of $\mathcal{S}_{{\sigma}}$. This requires that for any state $x_k \in \mathcal{S}_{{\sigma}}$, there exists a control $u_k$ such that $V_{{\sigma}}(x_{k+1}) \ge 0$. 
{For safety-critical AVs, a natural unifying viewpoint is to encode
safety through a scalar mode-conditioned safety value function
\(
V_{\bar{\sigma}}:
\mathcal{X}
\rightarrow
\mathbb{R}.
\)
For a fixed mode $\bar{\sigma}$, define its zero-superlevel set as
\(
\Omega_{\bar{\sigma}}^{V}
:=
\left\{
x\in\mathcal{X}
\;\middle|\;
V_{\bar{\sigma}}(x)\geq 0
\right\}
\subseteq
\mathcal{X}_{\mathrm{safe},\bar{\sigma}}.
\)
The set $\Omega_{\bar{\sigma}}^{V}$ defines the certified region of
safe operation associated with the value function. The core mechanism
of the filter is to enforce the forward invariance of
$\Omega_{\bar{\sigma}}^{V}$. Thus, for every
$x_k\in\Omega_{\bar{\sigma}}^{V}$, the filter must select a control
$u_k$ such that
\(
V_{\bar{\sigma}}(x_{k+1})\geq 0.
\)}

Contingency events, such as cut-ins or the appearance of occluded traffic, are modeled as discrete mode switches in the hybrid system model \eqref{eq:hybrid_system}. These switches can induce instantaneous discontinuities in the safety landscape, drastically altering the feasible control space.\footnote{
% Mathematically, a state $x_k$ that is safe under the current mode $\sigma$ (i.e., $V_{\sigma}(x_k) \ge 0$) may immediately become unsafe under the new mode $\sigma'$ (i.e., $V_{\sigma'}(x_k) < 0$) before any continuous state evolution occurs. 
{Mathematically, a state $x_k$ that is certified safe under a fixed
current mode $\bar{\sigma}$, meaning
\(
V_{\bar{\sigma}}(x_k)\geq 0,
\)
may lie outside the certified safe region immediately after a transition to a different mode
$\bar{\sigma}'$, meaning
\(
V_{\bar{\sigma}'}(x_k)<0,
\)
even though the continuous state has not yet evolved.}
Conventional continuous-time safety certificates do not natively guarantee invariance across such arbitrary discrete jumps without robust pre-computation. As a result, the reactive supervisor must react immediately upon detecting a mode change, updating its control decision to account for the new safety landscape. Two prominent methods that adopt this reactive constraint-enforcement viewpoint are control barrier functions (CBFs)~\citep{ames2019control} and Hamilton-Jacobi (HJ) reachability analysis~\citep{choi2021robust}.}
 
 However, it is crucial to note that invariance is not guaranteed if the transition occurs in an inevitable collision state (e.g., a cut-in that is "too fast to stop"). In such cases, the supervisor may need to intervene preemptively—before the mode switch physically manifests—if the state approaches the boundary of the safe set for a predicted future mode. \footnote{This necessitates coupling the supervisor with a predictive module, bridging the gap to proactive paradigms discussed in Sec.~\ref{sec:proactive}.}  

% CBFs, which can be viewed as special instantiations of safety value functions,  offer a geometric approach to enforce safety by establishing the forward invariance of a {safe set $\mathcal{C}_\sigma \subseteq \mathcal{S}_\sigma$}~\citep{ames2019control}. This is enforced by constraining the time derivative of the safety function $V_{{\sigma}}(x)$ (typically denoted $h_{{\sigma}}(x)$ in CBFs). 
CBFs, which can be viewed as special instantiations of safety value functions,  offer a geometric mechanism for enforcing safety through  the forward invariance of {a mode-conditioned set
\[
\mathcal{C}_{\bar{\sigma}}
:=
\left\{
x\in\mathcal{X}
\;\middle|\;
B_{\bar{\sigma}}(x)\geq 0
\right\}
\subseteq
% \Omega_{\bar{\sigma}}^{V}. 
\mathcal X_{\mathrm{safe},\bar\sigma}.
\]
This invariance condition is enforced by constraining the evolution of
the barrier function $B_{\bar{\sigma}}(x)$ under the
mode-conditioned dynamics $f_{\bar{\sigma}}$.}
{The typically handcrafted and low-dimensional parameterization of CBFs provides a significant advantage}: it enables explicit, interpretable safety specifications that encode domain knowledge and {can
establish verifiable forward-invariance guarantees under valid dynamics,
disturbance bounds, and state information.}

% {However, this design flexibility introduces fundamental challenges. Constructing a valid CBF that guarantees persistent feasibility under system dynamics and input constraints remains difficult for complex driving scenarios~\citep{black2023future,black2023safety,deshpande2025energy}.} An invalid CBF can lead to optimization infeasibility during critical contingencies (e.g., aggressive cut-ins), rendering the safe control set empty~\citep{wabersich2023data}. Moreover, even when valid, CBFs typically yield conservative safe sets that are subsets of the maximal control-invariant set. This inherent conservatism restricts the system from operating within a reduced safe region, potentially limiting operational efficiency. 
{
However, this design flexibility also introduces fundamental challenges. Constructing a valid CBF that guarantees persistent feasibility under system dynamics and input constraints remains difficult in complex driving scenarios~\citep{black2023future,black2023safety,deshpande2025energy}.
}
An invalid CBF can lead to optimization infeasibility during critical contingencies, such as aggressive cut-ins, rendering the admissible safe-control set empty~\citep{wabersich2023data}.
{
To address this feasibility issue, \citet{chen2021backup} proposed Backup CBFs, which use a predefined backup policy to implicitly construct a recoverable control-invariant set, thereby guaranteeing the feasibility of the associated online safety-filter optimization problem under the stated assumptions \citep{kim2026backup}. Nevertheless, the recoverable set induced by a fixed backup policy generally remains an inner approximation of the maximal control-invariant set. This inherent conservatism restricts the system to a reduced safe operating region, potentially limiting operational efficiency.
}
 
On the other hand,  HJ reachability formulates the reachability of a target set as an optimal control problem.  In this way, the safe value function $V_{{\sigma}}(x)$ is obtained by solving a Hamilton-Jacobi-Isaacs partial differential equation (PDE), explicitly accounting for worst-case disturbances or adversarial interactions~\citep{bansal2017hamilton,fisac2019general}. 
% The zero-superlevel set of the viscosity solution typically recovers the maximal control-invariant set $\mathcal{S}_{\infty}$, providing the least conservative formal guarantee. 
{For a fixed mode $\bar{\sigma}$, the zero-superlevel set of the HJ
viscosity solution typically characterizes the maximal
control-invariant safe set, which we denote by
\(
\mathcal{X}_{\mathrm{inv},\bar{\sigma}}^{\max}.
\)}
However, exact computation of this PDE suffers from the "curse of dimensionality", rendering it intractable for high-dimensional AV states (e.g., $\geq$ 5D) without strong decoupling assumptions. Practical implementations typically resort to low-dimensional abstractions, decoupling assumptions, and offline precomputation, which can limit applicability in complex contingency driving scenarios.
Moreover, the resulting safe control policy tends to be overly conservative, leading to defensive driving behaviors even in nominal scenarios. These limitations have motivated recent efforts to synthesize more tractable and less conservative value functions, such as the Control Barrier-Value Function (CBVF)~\citep{choi2021robust}. The CBVF is the viscosity solution to a modified Hamilton-Jacobi-Isaacs variational inequality that introduces a discount rate $\gamma \geq 0$, structurally embedding a CBF-like decay constraint within the reachability formulation. This yields the maximal safe set while enabling a less conservative optimal policy that permits approaching the safety boundary at a controlled rate $\gamma$.
{Although this formulation yields a less conservative policy, CBVF typically relies on grid-based PDE solvers and thus suffers from the same scalability issues as standard HJ. To address this computational bottleneck, recent research has pivoted toward data-driven approximations reviewed in \citep{dawson2023safe,wabersich2023data}.} 

% {
% Accordingly, classical grid-based HJ reachability is primarily practical as an offline synthesis tool for low-dimensional models, although the resulting precomputed value function can subsequently support low-latency online safety evaluation.
% }

% \textbf{Connections:} {The safe set induced by} a valid CBF ($\mathcal{C}_{{\sigma}}$)  is strictly a subset of the maximal control-invariant set ($\mathcal{S}_{\infty}$) derived from HJ analysis ($\mathcal{C}_{{\sigma}} \subseteq \mathcal{S}_{\infty}$). Consequently, a valid CBF can be viewed as a computationally tractable but conservative lower bound on the optimal safety value function. 
{\textbf{Connections:}
A valid CBF induces a control-invariant inner approximation of the
maximal HJ safe set:
\[
\mathcal{C}_{\bar{\sigma}}
\subseteq
\mathcal{X}_{\mathrm{inv},\bar{\sigma}}^{\max}
\subseteq
\mathcal{X}_{\mathrm{safe},\bar{\sigma}}.
\]
The first inclusion need not be strict in every problem. Thus, a valid
CBF should be interpreted as a computationally tractable, potentially
conservative inner approximation of the maximal control-invariant set,
rather than as a lower bound on the value function itself.}
Practically, a significant research trend focuses on the learning-based convergence of these paradigms to mitigate their respective limitations—specifically, the manual design complexity of valid  CBFs and the curse of dimensionality inherent to HJ PDEs. Recent works leverage neural networks as universal function approximators to synthesize high-dimensional safety certificates. For instance, physics-informed frameworks such as DeepReach~\citep{bansal2021deepreach} learn a neural value function $V_{\theta}(x)$ that approximates the HJ PDE solution, whereas neural CBFs optimize network parameters to enlarge the safe set volume subject to validity constraints~\citep{dawson2023safe,liu2022safe2}. 
{To mitigate the black-box nature of the learned value functions, recent work introduces formal verification to prove the correctness of these functions~\citep{yang2025scalable}. 
Although this synthesis effectively shifts the computational burden offline, formally certifying that these learned representations provide rigorous safety guarantees for contingency events in safety-critical AVs remains a significant open problem. For a broader treatment of safe control methods, see related surveys~\citep{hsu2024safety,wabersich2023data,dawson2023safe}.  }

\subsubsection{Implementation Strategies}
Building on these principles,  implementation strategies generally fall into two categories based on how they modulate the nominal policy: optimization-based filtering (minimally invasive) and policy switching. 
% Both designs fit a unified supervisory architecture:   a logical contingency predicate is monitored, and when it is triggered, the supervisor ensures that the system remains inside a zero-superlevel safe set of $V_{{\sigma}}(x)$.
{Both designs fit a unified supervisory architecture: a logical
contingency predicate is monitored, and when it is triggered, the
supervisor enforces invariance of the mode-conditioned certified set
$\Omega_{\bar{\sigma}}^{V} 
\subseteq
\mathcal X_{\mathrm{safe},\bar{\sigma}}$.}

\textbf{Minimally Invasive Filters:}
These mechanisms actively modulate the nominal control input to satisfy safety constraints while minimizing deviation from the intended performance. This is most commonly instantiated as a CBF-based safety filter
that constrains the one-step evolution of the barrier function.
% \begin{equation}
% u_{\mathrm{safe}} = \arg\min_{u \in \mathcal{U}} ||u - \pi^{\mathrm{nom}}(x)||^2 \quad \text{s.t. } \dot{h}_{{\sigma}}(x, u) \geq -\alpha(h_{{\sigma}}(x)),
% \label{eq:cbf_qp}
% \end{equation}
% where $\dot{h}_{{\sigma}}(x, u) = \nabla h_{{\sigma}}(x) \cdot f_{{\sigma}}(x,u)$ represents the evolution of the safety margin ({where $\sigma$ remains fixed during the continuous interval between discrete transitions}), and $\alpha$ is a class-$\mathcal{K}$ function tuning the allowable approach rate to the boundary. 
{
For a fixed mode $\bar{\sigma}$ over the current state transition, a
discrete-time CBF filter can be written as
\begin{equation}
\begin{aligned}
u_{\mathrm{safe},k}
=
&\arg\min_{u\in\mathcal{U}}
\left\|
u-\pi^{\mathrm{nom}}(\mathcal{I}_k)
\right\|^2 \\
&\text{s.t.}\quad
B_{\bar{\sigma}}
\left(
f_{\bar{\sigma}}(x_k,u)
\right)
\geq
(1-\gamma)
B_{\bar{\sigma}}(x_k),
\end{aligned}
\label{eq:cbf_filter}
\end{equation}
where  $\gamma\in(0,1]$ regulates the allowable approach rate to the set boundary. 
When the filter is feasible, this condition ensures that
$x_{k+1}=f_{\bar{\sigma}}(x_k,u_{\mathrm{safe},k})$ remains in the
zero-superlevel set
$\mathcal{C}_{\bar{\sigma}}
=
\{x\in\mathcal{X}\mid B_{\bar{\sigma}}(x)\geq0\}$
whenever $x_k\in\mathcal{C}_{\bar{\sigma}}$, thereby establishing
forward invariance under the fixed mode.
When the resulting barrier constraint is affine in the control input,
the safety filter can be implemented as a quadratic program (QP).
}

% {For a fixed mode $\bar{\sigma}$, a continuous-time CBF filter can be
% written as
% \begin{subequations}
% \begin{align}
% u_{\mathrm{safe}}
% =
% &\arg\min_{u\in\mathcal{U}}
% \left\|
% u-\pi^{\mathrm{nom}}(x)
% \right\|^2 \\
%   & \text{s.t.}
% \quad
% \dot{B}_{\bar{\sigma}}(x,u)
% \geq
% -\alpha
% \big(
% B_{\bar{\sigma}}(x)
% \big),
% \label{eq:cbf_qp}
% \end{align}
% \end{subequations}
% where
% \[
% \dot{B}_{\bar{\sigma}}(x,u)
% =
% \nabla B_{\bar{\sigma}}(x)^{\top}
% f_{\bar{\sigma}}(x,u)
% \]
% denotes the evolution of the safety margin under the fixed mode
% $\bar{\sigma}$, and $\alpha$ is a class-$\mathcal{K}$ function that
% controls the allowable approach rate to the set boundary.}
% In discrete-time AV implementations, this is typically enforced as a discrete barrier condition $B_{{\sigma}}(x_{k+1}) \ge (1-\gamma)B_{{\sigma}}(x_k)$ with $\gamma \in (0,1]$, ensuring asymptotic convergence to or invariance of the safe set. 

\textbf{Switching Filters:}  Alternatively, supervision can be implemented as an explicit policy switch. Consistent with the value-function viewpoint, the control law is:
% \begin{equation}
% u_k \;=\; 
% \begin{cases}
% \pi^{\mathrm{nom}}(x_k), & \text{if } V_{\sigma_k}(x_k) \geq \delta,\\[2mm]
% \pi^{\mathrm{safe}}(x_k), & \text{otherwise},
% \end{cases}
% \label{eq:switching_filter}
% \end{equation} where $\delta > 0$ serves as a buffered safety margin. 
{\begin{equation}
u_k =
\begin{cases}
\pi^{\mathrm{nom}}(\mathcal{I}_k),
&
\text{if }
V_{\sigma_k}(x_k)\geq\delta,\\[2mm]
\pi_{\sigma_k}^{\mathrm{safe}}(x_k),
&
\text{otherwise},
\end{cases}
\label{eq:switching_filter}
\end{equation}
where $\delta>0$ is a buffered safety margin and
$\pi_{\sigma_k}^{\mathrm{safe}}$ denotes the backup policy associated
with the currently resolved mode $\sigma_k$.}
The supervisor permits the nominal policy while the state remains in the interior of the mode-specific invariant set and triggers the mode-conditioned backup policy $\pi^{\mathrm{safe}}$ (e.g., emergency braking) as the state approaches the boundary.

In practice, advanced AV safety architectures typically combine these patterns, prioritizing minimally invasive filtering for local deviations while reserving discrete mode switching for critical contingencies where safety constraints render the nominal optimization infeasible.
 
\subsubsection{Concrete Examples in AV Contingency Scenarios}
This section maps common AV contingencies to the unified hybrid formulation established in Section~\ref{sec:problem_formulation}, categorizing them by the specific system component that undergoes a discrete shift.

{The most fundamental class of contingencies involves discrete shifts in the system's dynamics $f_{\sigma}$ (Eq.~\eqref{eq:cont_dyn}), arising from physical faults or controller failures.} 
 To address physical component failures, \citet{zhang2025safe} developed Fault-Tolerant CBFs (FT-CBFs) for nonlinear systems to guarantee safety despite actuator failures. They validated this framework on a Boeing 747 lateral control system, demonstrating that flight safety can be maintained even under severe rudder servo failures.
 For broader operational faults, \citet{singletary2022onboard} addressed the contingency of radio communication failure. Here, the loss of the pilot's link represents a discrete transition to a communication-loss mode; the onboard supervisor minimally modifies the pilot's last known commands to strictly enforce geofence boundaries during this blind phase. Similarly, when the nominal planner is a learning agent (e.g., end-to-end RL) or a fallible human operator, a policy failure manifests as a dynamics mismatch contingency—where the actual closed-loop evolution diverges from the supervisor's nominal model. {\citet{fisac2019general} developed a general HJ reachability framework
in which a value function characterizes the set of states from which constraint violations are avoidable. This framework can override a learning controller when dynamics mismatches are detected, as demonstrated in an aerial vehicle tracking task under sudden wind disturbances.}  Extending this to human-in-the-loop systems, \citet{oh2025safety} proposed a Human-Centered Safety Filter (HCSF) for high-speed racing vehicles {to handle the contingency of} a human driver losing control (e.g., missing a braking point). While effective for safeguarding black-box policies, these methods suffer from a fundamental model-mismatch issue: the shield’s safety guarantees are only as reliable as its underlying model~\citep{lu2025safe,goodall2023approximate,odriozola2023fear}. Safety can be compromised whenever a contingency (e.g., an icy road) is unmodeled in both the nominal agent and the shield. Furthermore, relying on supervisors for these internal failures can induce "skill degradation", as noted by \citet{oh2025safety}, where the nominal agent fails to learn robust behaviors because the filter systematically masks its errors.

% In contrast to internal dynamics faults or nominal controller failures, environmental contingencies are governed by the logical mode $\sigma$ in Eq.~\eqref{eq:mode_logic}. In this context, a mode switch does not alter the physical capability of the AV but fundamentally contracts the admissible safe set $\mathcal{X}_{\mathrm{safe},\sigma}$ 
{In contrast to internal dynamics faults or nominal-controller failures,
many environmental contingencies primarily alter the logical mode in
\eqref{eq:mode_logic}. For a fixed resolved logical mode
$\bar{\sigma}$, such a transition may leave the physical dynamics
unchanged while modifying the admissible set
$\mathcal{X}_{\mathrm{safe},\bar{\sigma}}$, for example, when a newly
detected obstacle or a resolved agent intent introduces additional
collision-avoidance constraints~\citep{qian2025collision}.}  For unexpected spatial constraints, \citet{strasser2024collision} applied a minimalist intervention, velocity saturation, to autonomous e-scooters, clamping the speed command when a sudden obstacle contracts the available braking envelope.  {For AV navigation in unknown environments, \citet{bajcsy2019efficient} addressed the contingency of sudden detection of a previously occluded obstacle, triggering a switch to a conservative safe kernel based on HJ analysis.  Moreover, \citet{agrawal2024gatekeeper} considered an autonomous quadrotor navigating an initially unknown obstacle environment, where the proposed framework uses online safety verification to ensure that only trajectories retaining a certified backup maneuver are executed.}
To further account for active agents, \citet{liu2016enabling} actively tracked the intention of other road participants and used an idea similar to CBF-QP to address contingencies (e.g., cut-in) in freeway driving;  \citet{leung2020infusing} applied HJ reachability to autonomous traffic weaving, treating any human action that eliminates a collision-free "escape maneuver" as a contingency. Similarly, \citet{hu2023deception} addressed "deception games", {where the contingency is a mismatch between the AV's belief and the human's actual hidden intent. To safeguard learning agents against these external hazards, \cite{wang2024safe} and \citet{raeesi2025safe} shield RL agents by constructing reachable sets that veto unsafe actions, ensuring safe driving within unpredictable traffic flows.}  While worst-case formulations like \cite{leung2020infusing} provide rigorous guarantees, they can induce the "Frozen Robot Problem"~\citep{trautman2010unfreezing}. 
% Probabilistic relaxations \cite{hu2024active} mitigate this by intervening only when risk exceeds a threshold, but they introduce a validity gap if the underlying belief model is miscalibrated. 

{Another challenging class concerns the degraded observability under~\eqref{eq:observation}, which makes the ego state uncertain.}
% \textcolor{green}{Addressing sensor corruption, \citet{zhang2025safe} proposed integrating a bank of state estimators directly into the barrier function constraints, allowing the control system to maintain safety properties even under active sensor attacks or faults.} 
For example, \citet{chen2021safe} used approximate reachability to handle ego-vision errors that corrupt safety boundary estimations in autonomous racing. Addressing direct hardware faults, \citet{laine2020eyesclosed} proposed an "Eyes-Closed Safety Kernel" for visual-inertial navigation, which switches to a proprioceptive (IMU)-only fallback policy when vision fails. However, these approaches typically rely on idealized fault detection, assuming the system can instantly identify the onset of a sensor fault or occlusion. In practice, faults are typically gradual or ambiguous (e.g., calibration drift), leading to reactive latency or chattering between modes. Moreover, binary fallback policies tend to be excessively conservative for highway speeds. Moving beyond binary switching, \citet{yun2025atom} introduced an adaptive CBF (ATOM-CBF) that continuously adjusts the safety margin based on real-time epistemic uncertainty. This allows the filter to gracefully degrade performance in the presence of out-of-distribution (OOD) measurements, maintaining safety without triggering abrupt maneuvers. While this mitigates the latency of binary detection, it introduces a critical dependency on uncertainty calibration. If the learned uncertainty estimate is miscalibrated (e.g., underestimating the risk of an OOD sample), the resulting safety margin will be insufficient to guarantee invariance.

Across these examples, task-preserving runtime safety filters are typically analyzed under simplified dynamics and idealized contingency predicates provided by upstream modules. However, safety filters must interface with perception and prediction systems that expose complex, learned uncertainties in deployed AV stacks. Additionally, they must also operate under tight real-time and hardware constraints. Systematically quantifying how these filter guarantees compose with upstream perception uncertainty remains a critical open challenge.

\subsection{Task-Terminating Fail-Safe Supervision}\label{subsec:fail_safe} 
While task-preserving filters enforce stepwise safety during nominal operation, contingencies such as severe sensor degradation, mechanical failure, or unresolvable environmental conflicts may render the mission infeasible. In such cases, the system must execute a deliberate transition to a safe terminal state. This task-terminating regime is the domain of fail-safe supervision, a specialized safety filter typically residing at the highest level of the control hierarchy.

Fail-safe supervision prescribes fallback maneuvers and specifies their triggering conditions, with the primary objective of achieving a \emph{Minimal Risk Condition} (MRC), such as safely stopping on the road shoulder~\citep{stolte2021taxonomy,iso4804_2020}.
The mathematical objective shifts from \emph{invariance} to \emph{convergence}. 
% Unlike task-preserving methods which maintain $x_t \in \mathcal{X}_{\text{inv}}$, fail-safe supervision seeks a policy $\pi^{\mathrm{fs}}$ that ensures finite-time convergence to a terminal safe set $\mathcal{X}_{\text{MRC}} \subset \mathcal{X}_{\text{safe}}$ while maintaining safety constraints throughout the transition. In practice, this terminal set $\mathcal{X}_{\text{MRC}}$ typically corresponds to a static "stop" state (i.e., zero velocity), ensuring the vehicle remains passively safe indefinitely.

% Formally, the set of valid fail-safe policies is defined as:
% \begin{align}
%     \pi^{\mathrm{fs}} := \Big\{ \pi \ \Big|\ & \exists T < \infty : \forall t \in [0, T],\ \\ \notag
%     &x_t \in \mathcal{X}_{\mathrm{safe}} \land \forall t \ge T,\ x_t \in \mathcal{X}_{\mathrm{MRC}} \Big\}. 
% \end{align}
{Unlike task-preserving methods, which maintain the state within
$\mathcal{X}_{\mathrm{inv},\bar{\sigma}}$}, fail-safe supervision seeks
a policy that guarantees finite-time convergence to a
mode-dependent minimal-risk-condition set {$\mathcal{X}_{\mathrm{MRC},\bar{\sigma}}
\subseteq
\mathcal{X}_{\mathrm{safe},\bar{\sigma}}$,
while satisfying the mode-dependent safety constraints throughout the
transition. In practice,
$\mathcal{X}_{\mathrm{MRC},\bar{\sigma}}$ typically corresponds to a
passively safe stopped condition, although its precise definition may
depend on the contingency mode.}

{Formally, the set of valid fail-safe policies for mode
$\bar{\sigma}$ is
% \begin{equation}
% \Pi_{\mathrm{fs},\bar{\sigma}}
% :=
% \left\{
% \pi
% \;\middle|\;
% \begin{array}{l}
% \exists T<\infty
% \text{ such that }
% x_{\tau}
% \in
% \mathcal{X}_{\mathrm{safe},\bar{\sigma}},
% \quad
% \forall \tau\in[0,T],
% \\[1mm]
% x_{\tau}
% \in
% \mathcal{X}_{\mathrm{MRC},\bar{\sigma}},
% \quad
% \forall \tau\geq T
% \end{array}
% \right\}.
% \label{eq:failsafe_policy_set}
% \end{equation}} 
{\begin{equation}
\hspace{-2mm}\Pi_{\mathrm{fs},\bar{\sigma}}
:=
\left\{
\pi \,\middle|\,
\begin{aligned}
&\exists K<\infty:\;
x_k\in\mathcal{X}_{\mathrm{safe},\bar{\sigma}},
&& \quad 0\leq k\leq K\\
&x_k\in\mathcal{X}_{\mathrm{MRC},\bar{\sigma}},
&& \forall k\geq K
\end{aligned}
\right\},
\label{eq:failsafe_policy_set}
\end{equation}} where $\mathcal{X}_{\mathrm{MRC},\bar{\sigma}}$ is assumed to be
control invariant under an associated terminal policy.
} 

A primary challenge in this domain is the online synthesis of verified, dynamically feasible, safe backup trajectories. We categorize existing approaches by their abstraction level: continuous trajectory synthesis, discrete logical supervision, and system-level safety architectures.

% The primary algorithmic challenge is synthesizing a dynamically feasible trajectory that terminates in $\mathcal{X}_{\text{MRC}}$.
{The primary algorithmic challenge is synthesizing a dynamically
feasible trajectory that terminates in the mode-dependent set
$\mathcal{X}_{\mathrm{MRC},\bar{\sigma}}$ while remaining within
$\mathcal{X}_{\mathrm{safe},\bar{\sigma}}$ throughout the transition.}
Early approaches relied on infinite-horizon invariant sets~\citep{blanchini1999set}, which are typically computationally prohibitive for real-time applications.
To address this, \citet{Pek2018SafeStates} proposed under-approximating invariant sets online.  Building on this, \citet{Pek2021Failsafe} introduced an online verification framework based on convex optimization. By verifying that a valid trajectory to a safe state always exists within a finite horizon, this method ensures recursive feasibility without computing the full maximal invariant set.
Safe stochastic MPC has extended these ideas by explicitly planning safe backup trajectories, accounting for operational uncertainties through set-based reachability analysis~(\cite{Brudigam2023}). 
However, recursive feasibility in these frameworks  relies on the assumption that surrounding agents adhere to modeled behaviors (e.g., traffic rules).  This limits robustness in non-compliant scenarios where the fail-safe maneuver itself might provoke a collision.

For discrete system failures (e.g., GNSS signal loss or software process crashes), fail-safe logic is typically synthesized using formal methods~\citep{schurmann2017ensuring, Krook2019, krook2020formal}.  For instance, when considering ego-vehicle localization failures, such as GNSS faults, specifications formalized in linear temporal logic (LTL) have been used to verify a supervisory controller that manages scenario switches between nominal planners and safe-stop trajectory planners~\citep{Krook2019}. Building on this, automated synthesis techniques have been explored. \cite{ krook2020formal} utilized supervisory control theory and reactive synthesis to generate correct-by-construction tactical planners.  By shifting the effort from manual implementation and verification to formal requirement specification, these methods generate correct-by-construction supervisory controllers that guarantee the system transitions to the correct fallback mode upon fault detection.  A fundamental limitation here is the abstraction gap. Formal synthesis typically operates on a discrete abstraction of the system. Mapping these discrete guarantees to the continuous, nonlinear dynamics of a vehicle executing a high-speed emergency stop remains a significant validation challenge.

Complementing algorithmic supervision, system-level research focuses on hardware and software architectures with inherent fault tolerance~\citep{julitz2023computer, grubmuller2019fault}. For example, \cite{julitz2023computer} proposed fault-tolerant hardware architectures for AVs, emphasizing redundancy, diversity, separation, self-diagnosis, and reconfiguration to enhance system reliability.
 Real-world implementations, such as the Mercedes-Benz DRIVE PILOT, integrate redundancy in braking and steering to ensure controllability during component outages~\citep{mercedesbenz2022}. While redundancy provides the highest robustness, it incurs high cost and weight penalties. Moreover, managing the "handover" logic between redundant systems (e.g., voting schemes) introduces its own complexity and potential failure modes, which are typically under-represented in algorithmic safety literature.

\subsubsection{Concrete Examples in AV Contingency Scenarios}
In practice, the selection of a fail-safe strategy is driven by the specific nature of the contingency event.  

\textbf{Actuation and Dynamics Faults:} 
For critical hardware failures, research concentrates on fault-mitigating control algorithms and redundant system architectures designed to preserve basic vehicle controllability for emergency maneuvers~\citep{yue2019automated, boudali2018emergency, khelladi2020emergency, lodder2023optimization, duerr2020realtime, li2023novel}. {For example, \citet{li2023novel, li2020shared} proposed a trust-based shared control framework to address tire blowouts on highways. This framework dynamically allocates steering authority based on the driver's panic level to prevent destabilizing inputs. For power steering failures, \citet{lodder2023optimization} utilized nonlinear MPC to compute safe-stop trajectories that explicitly account for limited steering torque. Expanding to simultaneous fault detection and control, \citet{lee2022adaptive} applied an adaptive sliding mode control (SMC) scheme to AVs. This approach enables real-time compensation for actuator loss of effectiveness while maintaining lateral stability.}

\textbf{Environmental and Interactive Hazards:}
For contingencies arising from interactive uncertainty (e.g., the unpredictable intentions of other agents), techniques leveraging reachability analysis and online verification are preferred to ensure robust, worst-case compliant maneuvers~\citep{Althoff2013, Magdici2016, schurmann2017ensuring, Pek2021Failsafe, Brudigam2023}. In conditional automation scenarios, the challenge extends to human-vehicle interaction. Addressing this, \citet{Xue2023Shared, Xue2022Override} developed shared control frameworks that estimate driver intention during critical failures. These systems utilize real-time risk assessment to determine the feasibility of a safe driver takeover, facilitating a manual transition only when the automated system's capabilities are exceeded.

\textbf{Observability and Perception Failures:} 
In contrast, sensor and localization faults, such as GNSS or LiDAR failures, require discrete logic to switch state estimators. The focus shifts to strategies for fault detection, sensor isolation, and fallback localization to maintain a reliable state estimate~\citep{Krook2019, Viana2022, grubmuller2019fault}. {Addressing extreme power blackouts where all external positioning is lost, \citet{jonasson2020} validated a "Blind Safe Stop" application relying exclusively on wheel-speed and pinion-angle sensors.} Similarly, perception system malfunctions are mitigated through methods that employ virtual sensors or revert to conservative, predefined environmental assumptions~\citep{Xue2019Virtual, Xue2018Fallback}. Specifically, these methods project phantom obstacles into the undetected area, forcing the fail-safe planner to execute conservative avoidance maneuvers as if the blind spot were occupied.

Despite these advances, fail-safe supervision remains an open research problem. Most methods operate under idealized assumptions regarding fault detection, actuation, and communication. Constructing backup trajectories robust to tracking errors, actuator saturation, and noisy fault indicators is critical. Furthermore, current architectures rarely link formal fail-safe concepts to regulatory definitions of MRCs or rigorously analyze the coupling between fail-safe actions and upstream perception modules. Bridging these gaps is essential for translating algorithmic prototypes into certifiable safety architectures.

\section{Proactive Safety Paradigms}\label{sec:proactive}   
The {Reactive Safety} paradigms discussed in Section~\ref{sec:reactive_safety} operate on the premise that the logical contingency mode is resolved or reliably estimated.  {Under valid modeling, timely mode detection, and online feasibility
assumptions, they can provide mode-conditioned safety guarantees, often
with relatively low online computational cost. However, their inability
to reason explicitly over unresolved future modes typically necessitates
worst-case assumptions, resulting in conservative behavior.}   For example, a false-positive pedestrian detection may trigger unnecessary braking, compromising task efficiency. 

This motivates Proactive contingency planning frameworks that explicitly reason about multiple possible futures rather than filtering a single plan post hoc.
In such frameworks, the evolution of logical modes is represented by a scenario tree.
A single non-anticipative policy is optimized over this tree, enforcing identical actions for scenarios that remain informationally indistinguishable.   

We formalize this perspective as a CB-DP, as introduced in Section~\ref{subsec:proactive_paradigm}. The framework provides an optimization-based approach for Proactive contingency planning, supporting predefined or dynamically generated branches. Crucially, it enables sequential branching decisions that defer final action commitments until key uncertainties are resolved, reducing conservatism while maintaining operational safety throughout.

\begin{figure} 
     \centering
     \begin{subfigure}[b]{0.78\linewidth}
         \centering
         \includegraphics[width=\linewidth]{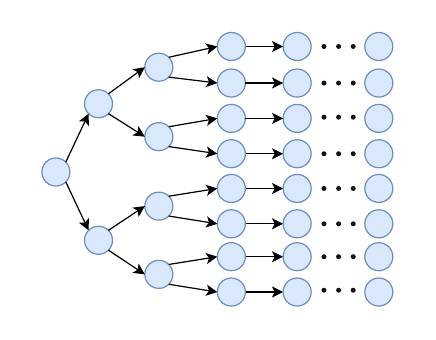}
         \caption{}
         \label{fig:tree_a}
     \end{subfigure}
     \centering
     \begin{subfigure}[b]{0.78\linewidth}
         \centering
         \includegraphics[width=\linewidth]{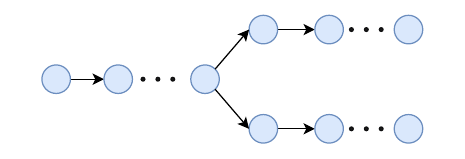}
         \caption{}
         \label{fig:tree_b}
     \end{subfigure}
      \centering
     \begin{subfigure}[b]{0.78\linewidth}
         \centering
         \includegraphics[width=\linewidth]{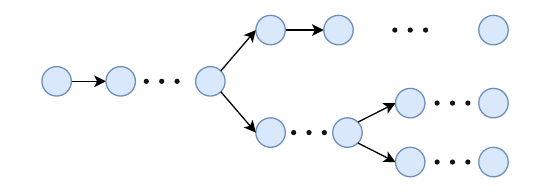}
         \caption{}
         \label{fig:tree_c}
     \end{subfigure}
        \caption{Representative tree structures.  (a): A tree structure commonly arises in min–max MPC formulations, where the tree branches in the early stages and stops branching after a finite number of stages to keep the tree size manageable. (b) and (c): In contingency MPC formulations, the tree typically exhibits various structures tailored to different applications.}
        \label{fig:tree_structure}
        \vspace{-0pt}
\end{figure} 

\subsection{From Min-Max MPC to Contingency MPC} 
\label{subsec:contingency_mpc} 
Min-max MPC has long served as a fundamental framework for robust control under bounded uncertainty. 
\cite{scokaert1998min, batkovic2021robust} distinguished between open-loop and closed-loop variants: the former generates a single control sequence that hedges against all admissible disturbance realizations, typically resulting in overly conservative behavior, whereas the latter leverages feedback to adapt actions in response to realized disturbances.
In classical formulations, uncertainty is represented as bounded disturbances acting on the system dynamics
\citep{scokaert1998min}.

Early open-loop min–max schemes optimize a fixed control sequence across all possible disturbance sequences,  disregarding the fact that a receding-horizon controller naturally re-optimizes as new measurements become available \citep{Zheng1993}. 
This mismatch typically leads to excessive conservatism and infeasibility in practice. 
% In contrast, feedback-aware min-max MPC \citep{scokaert1998min} minimizes the worst-case cost while enforcing robust state and input constraints.
% By employing a local stabilizing law \( u = Kx \), a robust positively invariant terminal set, and a quadratic terminal cost, 
% such formulations guarantee recursive feasibility and robust stability.  
% \luyao{Building on these classical formulations, \cite{lucia2013multi} proposed a multi-stage min–max MPC framework 
% to represent the evolution of uncertainty as a scenario tree.}
In contrast, feedback-aware closed-loop min-max MPC \citep{scokaert1998min, lucia2013multi} represents the evolution of uncertainty using a scenario tree, as illustrated in Fig.~\ref{fig:tree_a}. 
Each tree branch corresponds to a possible realization of future disturbances, 
and non-anticipativity constraints \citep{rockafellar1976nonanticipativity} ensure that decisions remain identical across branches until the uncertainties are resolved. 
This tree structure can handle multiple disturbance realizations within a single optimization problem. 
However, its exhaustive coverage of all admissible disturbance combinations leads to exponential growth in complexity.
To reduce the computational burden, as shown in Fig.~\ref{fig:tree_a}, the scenario tree stops branching after a certain time step.

Contingency MPC, a specialized form of scenario-tree-based MPC, has been applied to AV planning due to its capability to handle uncertainty and constraints. 
For example, an AV may experience a sudden loss of traction or the emergence of unexpected obstacles \citep{alsterda2021contingency}, or face interaction dilemmas, such as whether a nearby vehicle will yield \citep{zhan2016non}. 
In such settings, worst-case optimization across incompatible outcomes can hinder responsiveness and produce inconsistent plans \citep{elango2025deferred}. 

To address this limitation, contingency MPC explicitly plans over a finite set of plausible future scenarios while maintaining a feedback-aware receding-horizon structure \citep{alsterda2019contingency}.  
Rather than committing to a single trajectory, it constructs a compact scenario tree consisting of a shared trunk for immediate execution 
and multiple branch-specific tails as safe alternatives \citep{hardy2010contingency, geurts2025contingency}. 
% A central feature of this formulation is the non-anticipativity constraint~\eqref{eq:rcbdp_nonant}, 
% which ensures that decisions along the trunk remain identical across all branches until information reveals the active mode. 
This preserves decision consistency and feasibility while enabling timely adaptation to unfolding events.
In contrast to the scenario tree used in min-max MPC, the tree in contingency MPC, as illustrated in Figs.~\ref{fig:tree_b} and~\ref{fig:tree_c}, often exhibits various structures, which are customized to suit different applications.

\subsection{Contingency MPC and Applications} 
\label{subsec:contingency_mpc_applications}  
The foundational concepts of contingency MPC were established by early studies addressing well-specified contingency scenarios.  These studies introduced a formal "Plan B" concept for specific vehicle-state hazards, such as unexpected loss of road friction, which induces a switch in the system dynamics~\eqref{eq:cont_dyn} \citep{alsterda2019contingency, alsterda2021contingency}. Other studies have focused on uncertainty in logical modes, including the resolution of binary interaction dilemmas (e.g., "yield versus go" at intersections), and the development of optimization-based planners that handle mutually exclusive obstacle predictions by optimizing multiple contingency paths with a shared initial segment under probabilistic collision constraints \citep{hardy2013contingency}.
 
Subsequent approaches extended this multi-hypothesis principle to handle multimodal predictions of other agents \citep{RosoliaBMPC2023, bouzidi2024motion}. 
In these methods, the motion of the other agents, predicted by rule-based or learned models, is typically nonreactive. To better account for interactions, \cite{qiu2020latent} and \cite{hu2024active} incorporated interaction-aware models and belief propagation into a tree-structured planning framework, allowing the planner to automatically balance trajectory cost reduction with information gathering. 
In safety-critical domains such as autonomous driving, risk-aware approaches have also been proposed to manage low-probability, high-impact outcomes during scenarios such as merging, cut-ins, and highway driving. 
These methods address such challenges by incorporating risk-sensitive objectives, many of which are based on Conditional Value at Risk (CVaR), as demonstrated in approaches such as Branch MPC \citep{chen2022interactive}, MARC \citep{li2023marc}, and EraBMPC \citep{ZhangEraBMPC2024}. 
Other notable risk-aware frameworks include the interaction-aware branch MPC proposed by \cite{wang2023interactive}, which employs smooth sigmoid approximations of chance constraints to reduce conservatism while maintaining safety in discrete multimodal scenarios. 
Additionally, RACP \citep{mustafa2024racp} introduced custom probabilistic metrics based on the product of collision probability and severity, enabling a more nuanced risk assessment. 
% Such planning is typically coupled with high-level decision-making frameworks, such as Partially Observable Markov Decision Processes (POMDPs), to ensure the branching structure is guided by a robust, long-term belief state \citep{ulfsjoo2022integrating}.  
{
More generally, probabilistic prediction and belief models can guide
scenario construction, branch weights, and risk-sensitive objectives,
while the logical mode separately represents whether the relevant
behavior remains uncommitted or has been treated as committed
\citep{ulfsjoo2022integrating}.
}

Most existing contingency-planning schemes lack explicit occlusion assessment. 
To further account for potential phantom vehicles in occluded driving environments, an occlusion-aware contingency game was introduced within a receding-horizon planning framework \citep{qiu2024inferring}.  Alternatively, \citet{nyberg2025hope} integrated formal reachability analysis into a tree-based motion planner. By over-approximating the forward reachable sets of occluded agents, this framework guarantees the existence of a valid braking contingency without the excessive conservatism of assuming the object is always present.
Despite such advancements, scaling contingency planners to real-time operation in multi-vehicle settings remains challenging, especially in dense and partially observable traffic scenarios. {Distributed control architectures have been developed to improve
scalability in multi-vehicle systems~\citep{liang2023mas}. Within
contingency planning, Alternating Direction Method of Multipliers  (ADMM)-based decomposition can exploit scenario-tree
and multi-agent structure by coordinating parallel subproblems~\citep{huang2026fast}.} 
By iteratively solving decoupled subproblems, {ADMM can support parallel computation, although its runtime
and convergence depend on problem structure and penalty-parameter
selection~\citep{boyd2011distributed}}. For example, recent consensus ADMM-based formulations have generated safe contingency plans around occlusions
where multiple agents may appear unexpectedly in partially observed environments \citep{zheng2025safe, zheng2025oacp}.  

% As complexity grows, the need for formal safety guarantees becomes paramount.  
Formal safety guarantees for contingency MPC have been an active area of research.
A key research direction focuses on the integration of formal methods to provide verifiable safety against the worst-case outcomes of multimodal predictions. 
This includes the use of reachability analysis to certify that all planned branches are safe \citep{bouzidi2025reachability}, 
with some approaches leveraging online event-triggered learning for less conservative safety barriers while maintaining feasibility in interactive dense traffic \citep{yang2025safe}.  
To ensure persistent feasibility during planning, other methods used control-invariant sets $\mathcal{X}_{\mathrm{inv}}$, which guarantee that the vehicle can always be steered back into a precomputed safe state during aggressive maneuvers in mixed traffic (\cite{chen2023invariant, schweidel2022driver}). 
Building on this, \cite{geurts2025contingency} proposed a multi-horizon contingency MPC framework for safe learning that formally establishes robust recursive feasibility by employing control-invariant terminal sets, specifically using the union of two disjoint robust control-invariant sets for merging behind or ahead of other vehicles.  
Likewise, contingency-aware nonlinear MPC provided a unified formulation for complex driving maneuvers, achieving practical stability and recursive feasibility in dynamic traffic 
by enforcing control-invariant terminal sets with an LQR fallback \citep{lin2025contingency}.

While such algorithmic advances addressed the tractability of solving a given scenario tree, 
a key formulation challenge remains in determining the optimal branching point \citep{bouzidi2025reachability}. 
Fixing this branching point a priori, as is common, prevents adaptation to dynamic uncertainties 
and can induce either overly conservative behavior or unsafe reactive latency. 
Recent studies \citep{tas2018decision, bouzidi2025reachability} proposed adaptive strategies 
that determine the branching time online using information-based or reachability-based criteria. 
The adaptive‑branching mechanisms arising from these developments are further discussed in Section~\ref{subsec:dynamic_branching}.

\subsection{Game-Based Contingency Planning}  
While scenario-tree MPC handles discrete contingencies, game-based contingency planning enhances this capability by reasoning about strategic interactions among agents \citep{paden2016survey}.
Unlike approaches that treat other agents as passive participants following fixed predictions, game-theoretic planners assume that all agents are rational decision-makers optimizing their own objectives. This assumption enables the AV to anticipate and influence the behavior of other agents in interactive scenarios. Early work typically adopted Stackelberg leader-follower models, where the AV acts as the leader and optimizes its trajectory, assuming that others respond optimally as rational followers. While effective in generating socially aware behaviors, such models typically assume a single deterministic response from other agents, limiting their ability to capture multimodal interactions or prepare for contingencies.

To overcome this limitation, recent studies have integrated game-theoretic models into the framework of contingency planning, thereby explicitly accounting for multiple possible future scenarios. 
For instance, in complex negotiation scenarios such as lane merging, a matrix game can be formulated and solved to obtain multiple equilibria that capture the diverse driving decisions of other agents.
These equilibria define the branches of a scenario tree, allowing a branch MPC to generate contingency plans over plausible outcomes \citep{zhang2024automated}. 
A more integrated approach was introduced as contingency games for multi-agent interaction, which directly embeds multi-policy reasoning into the game problem \citep{peters2024contingency}.
In this framework, each agent selects from a discrete set of high-level policies (e.g., "yield" or "maintain speed"), and solves for a shared trunk trajectory that remains feasible and safe across all policy combinations.  
To facilitate high computational efficiency, recent work further modeled interactions as pairwise games, where the strategy space itself forms a trajectory tree, allowing the game solution to directly yield a branching contingency plan in real time \citep{ma2025trajectory}. 
Furthermore, \citet{huang2026fast} formulated interactive
trajectory planning under intent uncertainty as a Bayesian game.
Under the stated assumptions, its agent-form equivalent admits
a potential-game representation, enabling Bayesian Nash
equilibrium computation through a unified optimization problem.
To improve scalability, the authors employ a dual consensus ADMM algorithm for parallel optimization, making real-time interactive contingency planning feasible even in complex, multi-agent urban traffic scenarios.
 
Additionally, game theory can be leveraged to address contingencies arising from partial observability, such as the presence of agents in occluded driving areas \citep{Zhan-RSS-21}. 
By formulating the problem as a dynamic game with imperfect information, the planner can reason about the worst-case actions of a potential hidden agent in occluded urban interactions \citep{qiu2024inferring}.  
The resulting trajectory remains robust against the potential emergence of occluded agents, thereby intrinsically generating a contingency plan for high-risk occlusions. 

Despite their theoretical promise, game-based contingency methods face several practical challenges.
First, their performance is highly sensitive to the accuracy of the assumed or learned cost functions of other agents. 
Moreover, solving multi-agent dynamic games, especially those involving branching structures or imperfect information, remains computationally demanding, which constrains real-time applicability in dense traffic settings.  Critically,  these methods primarily address logical modes that alter the safety specification (e.g., intent ambiguity
that affects the admissible safe set), rather than modes governing internal system faults (e.g., degraded dynamics or sensor outages).

\subsection{{Learning-Based Contingency Planning}}  
Learning-based methods expand contingency planning beyond traditional analytical models, enabling AVs to predict, adapt, and react to open-world uncertainties using data-driven approaches. These methods generally fall into three categories: (i) behavioral topology and structured interaction modeling, (ii) multi-future prediction and contingency-policy learning, and (iii) foundation-model-driven anomaly detection and fallback planning.

Topology-aware representations provide a structural foundation for branching-based contingency planning by encoding the qualitative modes of multi-agent interaction to impose behavioral consistency. 
\citet{mavrogiannis2019multi,mavrogiannis2024abstracting}
use topological braids to represent equivalence classes of
joint trajectories in multi-agent navigation and road traffic,
capturing relative passing sides and crossing order.
While these studies focus on a single plan, they provide a foundation for representing interaction modes. 
Behavioral Topology (BeTop) \citep{liu2024reasoning} extended this topological modeling into a learning-based branching framework.  Rather than treating interaction as generic complexity, BeTop utilizes topological invariants to discretize multi-agent behaviors into a finite set of logical modes. In this framework, the contingency is formalized as the uncertainty over which topological class (e.g., a specific "yield" versus "cut-in" braid) will materialize.  By constructing a behavioral prior derived from braid theory, BeTop aligns prediction with planning via its BeTopNet, thereby improving coherence across interacting agents in dense traffic. Consequently, the branching structure of BeTopNet enables the planner to effectively manage the uncertainty inherent in multi-agent interactions. 
% Together, these approaches demonstrate how topology-aware frameworks improve consistency and realism in multi-agent contingency reasoning.

While topological methods define the abstract structure of interactions, the second category addresses contingency reasoning as a data-driven inference problem. Early sampling-based approaches, such as LookOut \citep{cui2021lookout}, introduced end-to-end multi-future prediction and planning frameworks that construct explicit contingency branches through stochastic rollouts and collision-set pruning.
However, this approach assumes that non-ego agents follow fixed, independent forecasts. 
To address this limitation, \citet{chen2023tree} proposed Tree-Structured Policy Planning, which combines the ego motion sampler and the ego-conditioned prediction model to generate multi-stage motion plans. 
\citet{Huang2024DTPP} advanced this line of work by unifying tree-structured planning and prediction through a query-centric Transformer model with learnable cost functions.
% \cite{qiu2020latent} further extended trajectory-tree branching to latent belief-space planning with belief-space DDP to handle latent uncertainties.
Subsequent works, such as {Contingencies from Observations} (\cite{rhinehart2021contingencies}), {Active Visual Planning} (\cite{packer2023anyone}), and {Conditional Behavior Prediction} (\cite{tolstaya2021identifying}), learn contingency-aware policies directly from raw observations, allowing adaptation to occluded or unobserved agents. 
% More recent frameworks, such as {MARC} (\cite{li2023marc}), {RACP} (\cite{mustafa2024racp}), and Non-conservative Planner (\cite{yang2025safe}), incorporate risk-aware optimization or online learning of reachable sets to balance safety and efficiency across uncertain futures. 
Leveraging generative models, CoPlanner \citep{zhong2025coplanner} advances this direction with a diffusion-based motion planner that jointly generates interactive multi-agent trajectories, enforcing a shared short-term segment for stability and diverse long-horizon branches for contingency. However, CoPlanner lacks recursive feasibility guarantees and relies on fixed branching and uniform scenario weighting, which limits adaptability to diverse driving contexts. These learning-based approaches move contingency planning from fixed scenario enumeration toward adaptive, probabilistic reasoning over learned future distributions.

When AVs encounter scenes that differ significantly from the training data, they may experience OOD semantic failures that traditional contingency planners do not anticipate. To address this, recent work uses the general knowledge of foundation models to interpret and react to unforeseen high-level hazards. \cite{sinha2024real} introduced a two-stage large language model (LLM)-based framework: a fast anomaly test in an LLM embedding space triggers a slower generative stage that decides how to respond, while the branch MPC maintains feasible fallback trajectories to ensure safety during the generative latency. 
The fast stage runs at real-time rates on embedded hardware, and the approach has been validated in driving scenarios under AV perception failures in CARLA, as well as on {quadrotor landing tasks.  
Moving from reactive fallback selection to proactive failure prevention, \cite{ganai2025real} proposed FORTRESS, a framework that uses multimodal foundation models to anticipate semantic failure modes and generate new, dynamically feasible fallback plans for AVs in real time. } This approach enhances robustness by synthesizing strategies adapted to unforeseen hazards (e.g., road closures, urban disruptions, and semantically unsafe regions) rather than relying on predefined sets.   These large-model-driven systems represent a significant evolution in contingency planning, combining low-level dynamic safety with high-level semantic reasoning to achieve more robust autonomy.

 \begin{table*}[t]
    \centering
    \scriptsize
    \renewcommand{\arraystretch}{1.4} 
    \setlength{\tabcolsep}{6pt} 
    \caption{{Comparative analysis of Reactive versus Proactive contingency planning paradigms.}}
    \label{tab:comparison}
    \begin{tabularx}{\textwidth}{l X X}
    \toprule 
    \rowcolor{arbeige} 
    \textbf{Criterion} & \textbf{Reactive Paradigm} (Section \ref{sec:reactive_safety}) & \textbf{Proactive Paradigm} (Section \ref{sec:proactive}) \\
    \midrule
    \textbf{Core Assumption} & 
    \textbf{Logical uncertainty is resolved.}  & 
    \textbf{Logical uncertainty is unresolved.} 
    \\   \midrule
    % \textbf{Safety Guarantee} & 
    % \textbf{State-Based Invariance or Convergence.} Guarantees state constraints ($x \in \mathcal{X}_{\text{inv}}$) or finite-time safe termination ($x_T \in \mathcal{X}_{\text{MRC}}$) \textcolor{red}{for all bounded disturbances}.  & 
    % \textbf{Sample-Path Safety.} Guarantees safety for the finite set of modeled scenarios. Subject to risk if the realized scenario lies outside the modeled tree support.
    % % Subject to probabilistic risk if the scenario tree does not cover the true uncertainty distribution. 
    % \\ 
    \textbf{Safety Assurance}
    &
    \textbf{Conditional Invariance or Convergence.}
    Forward invariance of
    $\mathcal{X}_{\mathrm{inv}}$
    or finite-time convergence to an MRC
    ($x_T \in \mathcal{X}_{\mathrm{MRC}}$)
    under valid model, disturbance-set,  and feasibility
    assumptions.
    &
    \textbf{Sample-Path Safety.}
   Safety constraints are enforced over modeled branches; unmodeled
realizations and post-horizon safety require additional coverage and
terminal-set assumptions.
    \\
    \midrule
    \textbf{Conservatism} & 
    \textbf{Method-dependent.}
Worst-case certificates and fixed backup policies can be conservative, whereas minimally invasive filters may remain close to the nominal controller.     & 
    \textbf{Reduced (Recourse-Aware).} Deferring mode-specific commitment can reduce conservatism, although broad scenario coverage or robust branch constraints may still produce cautious behavior. \\
    \midrule
    \textbf{Computational Cost} & 
    \textbf{Offline-Heavy (Synthesis) / Online-Light (Execution).} Relies on precomputed safety critics (e.g., value functions, invariant sets). Runtime is typically a fast algebraic check or QP. & 
    \textbf{Online-Heavy.} Solves large-scale optimization problems (e.g., scenario trees) at every timestep. Requires specialized solvers (e.g., ADMM, Riccati) for real-time performance. \\
    \midrule
    \textbf{Scalability} & 
    \textbf{Dimension-Limited.}
    Runtime execution scales well with agent count, but synthesizing a high-fidelity invariant set suffers from the "curse of dimensionality", limiting methods to low-dimensional systems.  & 
    \textbf{Scenario-Limited.} Complexity scales with the tree size, which depends on the branching factor and depth. Managing combinatorial uncertainty from multiple contingency events or long horizons requires aggressive pruning. \\
    \midrule
    \textbf{Adaptivity} & 
    \textbf{Triggered Response.} Intervenes only when the safety margin is depleted, or a fault is detected. Lacks the foresight to avoid the "point of no return." & 
    \textbf{Anticipatory Recourse.} Proactively preserves future options. Enables information-seeking actions to resolve uncertainty before branching. \\
    \bottomrule
    \end{tabularx}
\end{table*}  

\subsection{{Computational Methods}}
\label{subsec:computational_analysis}
The computational burden of contingency planning arises primarily from the dimensionality introduced by branching trajectories and the nonconvex constraints associated with safety-critical environments.

Trajectory planning problems are generally formulated as nonlinear optimization problems and solved using general-purpose solvers, such as IPOPT \citep{wachter_implementation_2006}, SNOPT \citep{gill_snopt_2005}, and Knitro \citep{pardalos_knitro_2006}. 
To efficiently solve the planning problem, a solver must leverage the sparsity pattern introduced by system dynamics.
In these general-purpose solvers, sparsity is typically detected and exploited by sparse linear solvers.
Alternatively, specialized optimal-control solvers and trajectory-optimization methods, such as acados \citep{Verschueren2021}, FORCESPRO \citep{FORCESNLP}, Crocoddyl \citep{mastalli20crocoddyl}, CFS \citep{liu2018convex}, and Aligator \citep{jallet2025}, exploit problem structure through methods including Riccati recursion, structured factorization, and sequential convex approximation. 
These solvers differ primarily in their approaches to constraint handling.   

Similarly, trajectory planning problems with tree structures can also be solved using the aforementioned general-purpose solvers or numerical optimal control solvers.  However, one additional computational challenge arises from the increased dimensionality of the decision variables due to the use of scenario trees. Therefore, a high-performance solver must leverage the sparsity pattern introduced by the scenario tree. 

Therefore, a high-performance solver must leverage the sparsity pattern introduced by the scenario tree. Yet, fully exploiting this sparsity may require dedicated solver routines or interfaces beyond those typically used for standard chain-structured optimal-control problems. This motivates the development of tailored Riccati recursion methods \citep{FRISON201714399} and matrix factorization techniques \citep{Klintberg2017, schwan2025piqp_multistage}.
% These customized methods typically involve performing multiple independent factorizations along the scenario tree paths, potentially in parallel, followed by an additional factorization to satisfy causality constraints near the tree root.
These customized methods exploit tree-structured sparsity through recursive or block-structured factorizations, with independent computations on separate branches or subtrees potentially performed in parallel.

{Another category of approaches applies the so-called dual decomposition technique to the scenario-tree problem \citep{Klintberg2016Dual, kouzoupis_dual_2019, kouzoupis_dual_2018}. The subproblems can be addressed in parallel and subsequently coordinated by solving a dual Newton system.}
Beyond these structure-specific techniques, consensus ADMM provides a general decomposition framework for efficient distributed computation \citep{boyd2011distributed,ghadimi2015optimal}. By reformulating the nonlinear program into a series of
low-dimensional subproblems with consensus constraints, it enables scalable decomposition in contingency planning \citep{phiquepal2021control,zheng2025safe,huang2026fast}. 
Recent studies \citep{zheng2025occlusion,zheng2025oacp} have demonstrated that such formulations can achieve real-time performance in dense, occlusion-aware driving scenarios on both simulation and hardware AV platforms, with runtimes on the order of tens of milliseconds. 
Nevertheless, the convergence speed remains sensitive to the tuning of penalty parameters, and inappropriate choices may increase iteration counts or degrade solution quality, which limits robustness in safety-critical applications.  {
From a deployment perspective, the principal scalability bottleneck is the size of the online scenario tree. Structure-exploiting solvers and parallel decomposition can support real-time execution for bounded or selectively pruned trees. However, long-horizon, densely coupled, or extensively branching formulations remain computationally restrictive. Solver acceleration therefore enlarges the tractable operating regime but does not eliminate the fundamental dependence on tree depth, branching factor, and interaction dimension.
}

\section{Discussion}\label{sec:discussion} 
In this section, we compare the two primary contingency planning paradigms: \textbf{Reactive} and \textbf{Proactive} planning. We first synthesize their shared scientific principles and core trade-offs regarding safety, efficiency, and scalability. Subsequently, we examine the critical challenge of dynamically determining the branching point within these frameworks.

\subsection{{Relationships and Connections}} \label{subsec:comparative_analysis}   
Reactive Safety methods condition their intervention on a
safety-relevant logical mode that has been observed or reliably
inferred, whereas Proactive Safety methods retain multiple unresolved
mode-conditioned futures. Reactive methods therefore depend primarily on timely mode resolution and the validity of their safety certificates or fallback policies, whereas Proactive methods depend on mode coverage, prediction quality, and computational tractability. Table~\ref{tab:comparison} summarizes these differences, emphasizing the factors that influence the choice between Reactive and Proactive approaches. 

% \textbf{Safety versus Conservatism}
% The most critical trade-off lies between the strength of the safety guarantee and behavioral conservatism.  
% {Certificate-based  Reactive methods can provide  strong mode-conditioned safety guarantees  under valid model, information, disturbance, and feasibility
% assumptions}  because they typically enforce invariance against the worst-case boundary of the control-invariant safe set (typically synthesized offline).
% However, this comes at the cost of high conservatism and sub-optimality.  Once triggered, a Reactive planner treats every potential hazard as an immediate threat, typically forcing the AVs to behave abruptly.  Additionally, by intervening only at the constraint boundary, these methods can induce control chattering—a high-frequency switching behavior that degrades passenger comfort and control smoothness.   Proactive methods can mitigate these issues by modeling future recourse—the ability to adapt in the future \citep{birge2011introduction}. By verifying safety for specific branches in a scenario tree, Proactive planners optimize a "shared trunk" trajectory in a receding horizon fashion. This look-ahead capability allows for smoother, more optimal maneuvers that defer the commitment to a specific mode until the branching time, leveraging online information to mitigate the sub-optimality of purely reactive interventions.

\textbf{Safety versus Conservatism}
The most critical trade-off lies between the strength of the safety
guarantee and behavioral conservatism.
{
Certificate-based Reactive methods can provide strong mode-conditioned
safety guarantees under valid model, information, disturbance, and
feasibility assumptions because they enforce forward invariance of a
certified safe set, which is typically synthesized offline.
}
However, {the resulting safety certificates may be
conservative, leading to suboptimal behavior.}
Once triggered, a Reactive planner
{may treat uncertain or insufficiently characterized
hazards conservatively, potentially causing abrupt vehicle responses.}
Additionally, {filters that intervene near the constraint
boundary may exhibit} control chattering---a high-frequency switching
behavior that degrades passenger comfort and control smoothness.

Proactive methods can mitigate these issues by modeling future
recourse---the ability to adapt in the future~\citep{birge2011introduction}.
By {enforcing safety constraints along the modeled
branches of a scenario tree}, Proactive planners optimize a "shared
trunk" trajectory in a receding-horizon fashion. This look-ahead
capability {can support smoother and less conservative
maneuvers by deferring commitment to a specific mode until the branching
time and exploiting information obtained online.}

\textbf{{Computation versus Scalability}}
{
Reactive and Proactive methods differ primarily in where the computational burden is incurred.} Reactive methods typically shift the complexity offline (e.g., synthesizing value functions, invariant sets,  or barrier certificates). This makes them highly scalable at runtime (e.g., QP or lookup), capable of filtering high-frequency control loops. {Their main limitation lies in the offline synthesis stage, where exact grid-based HJ reachability and high-dimensional invariant-set computation remain difficult to scale.}

In contrast, Proactive methods are {online-heavy and scenario-limited}. While this offers greater flexibility to handle changing environments, it introduces a bottleneck: the size of the scenario tree limits scalability. Handling dense traffic typically requires heuristic pruning, which forces a trade-off between tractability and the risk of discarding critical scenarios. {Accordingly, exact high-dimensional reachability formulations and large unpruned scenario trees remain primarily offline or computationally restricted.}

\textbf{Synthesis: The Hybrid Architecture}
This comparative analysis reinforces the "synergistic integration" outlined in Section~\ref{subsec:connections}. 
The high online computational cost and finite-horizon limitations of Proactive methods can be mitigated by imposing offline synthesized invariant sets as terminal constraints on scenario-tree branches. { By anchoring short-horizon branches in certified invariant terminal sets, recursive feasibility can be established under standard assumptions regarding model validity, terminal-set invariance, and online optimization feasibility, without requiring a computationally prohibitive deep scenario tree. This architecture thereby separates near-term strategic adaptation from long-term safety and stability.
}
Conversely, the conservatism inherent in Reactive methods is reduced by using offline optimal control computation to synthesize high-performance safety critics.  {
Exact HJ reachability~\citep{bansal2017hamilton} can characterize a maximal control-invariant safe
set under suitable assumptions, whereas learning-based MPC and
differential dynamic programming~\citep{fisac2019general, hsu2024safety} may provide tractable approximations or
performance-oriented safety critics.  Reactive filters can serve as a runtime "safety net" for Proactive planners, intervening when necessary to maintain certified safety conditions, provided that their own certification assumptions remain valid. Consequently, future robust architectures will likely layer these paradigms, utilizing a Proactive planner for low-frequency strategic guidance and a Reactive filter for high-frequency safety enforcement. }

\subsection{Paradigm Suitability and Methodological Maturity}\label{subsec:trends}
Beyond the theoretical trade-offs, the classification in Table~\ref{tab:contingency_comparison} reveals distinct structural trends governed by the nature of the contingency.
First, we observe a strong correlation between the contingency source and the preferred planning paradigm. 
Specifically, {internal contingencies, such as actuator faults or sensor failures, are predominantly addressed via Reactive methods.} This is attributable to the abrupt, binary nature of system faults; when a component fails, the immediate priority is stabilization or safe termination (i.e., reaching an MRC), a task well-suited to low-latency fail-safe supervision rather than strategic lookahead. Conversely, {external contingencies involving interactive agents are increasingly addressed by Proactive branching methods.} In these scenarios, uncertainty stems from the evolving intent of other agents. Proactive planning allows the AV to influence this evolution and maintain recourse, whereas the Reactive filter tends to induce "frozen" behavior by treating potential interactions as immediate worst-case hazards.

Second, there is a clear methodological divergence regarding safety assurance, which fundamentally reflects the trade-off between the rigor of guarantees and the restrictiveness of modeling assumptions. Contingencies rooted in physical dynamics, whether kinematic hazards or interactive agents, are largely addressed by methods with formal guarantees $[\mathcal{F}]$ (e.g., invariant sets). However, these rigorous guarantees are predicated on strict, often idealized assumptions, such as bounded disturbance sets and perfectly known differential equations, which can be brittle in unstructured real-world environments. 

In contrast, contingencies rooted in semantic understanding and OOD anomalies (e.g., encountering unrecognized hazards~\citep{sinha2024real}) rely heavily on empirical $[\mathcal{E}]$ or probabilistic $[\mathcal{P}]$ approaches. Although they generally lack formal proofs for unrestricted semantic errors, these methods operate under less restrictive environmental assumptions, facilitating easier transfer to open-world scenarios where reliable error
bounds may be unavailable. This dichotomy highlights a critical "Verification Gap": while the field has developed rigorous formalisms to guarantee the vehicle will not violate physical constraints, these methods are predominantly state-based, relying on the assumption of bounded estimation errors.

{Recent advances in formal methods have attempted to extend verification to the perception stack.} Notable frameworks such as VerifAI \citep{dreossi2019verifai} enable closed-loop verification of perception-driven systems, while specialized techniques have been developed to certify the robustness of neural classifiers (e.g., for traffic sign detection) \citep{weng2018towards, shi2020robustness} and, more recently, 6D pose estimation \citep{luo2025certifying}. However, these rigorous guarantees are predominantly predicated on bounded perturbation models, typically limited to defined $L_{p}$-norm pixel noise, convex approximations of semantic perturbations, or constrained geometric transformations. Consequently, strict safety guarantees are typically lost when accounting for the potentially unbounded and semantically complex perception errors inherent in open-world OOD anomalies, where the magnitude of distribution shifts cannot be predefined. 

Thus, the field currently lacks equivalent formal frameworks to guarantee the correct interpretation of complex semantic scenarios, necessitating the future development of verified runtime monitors for learning-based components.  

\begin{table*}[t]
    \centering 
    \scriptsize
    \renewcommand{\arraystretch}{1.} % Slightly increase row spacing
    \setlength{\tabcolsep}{3pt} %   
    \caption{{Classification of representative literature in contingency planning. Approaches are categorized by the nature of the contingency and the planning paradigm. \textbf{Guarantee Key:} $\mathcal{F}$ = Formal/Robust Invariance; $\mathcal{P}$ = Probabilistic/Stochastic; $\mathcal{E}$ = Empirical/Heuristic.}}      
    \label{tab:contingency_comparison}  
    % main_table.tex
\begin{tabularx}{\textwidth}{@{} p{3.2cm} X X @{}}
    \toprule
    \rowcolor{arbeige} 
    \textbf{Contingency Source} & \textbf{Reactive Approaches} (Sec. \ref{sec:reactive_safety}) & \textbf{Proactive Approaches} (Sec. \ref{sec:proactive}) \\ 
    \midrule

    %=================================================================
    % PART 1: EXTERNAL CONTINGENCIES
    %=================================================================
    \multicolumn{3}{@{}l}{\textbf{\color{arblue}I. External Contingencies (Environmental \& Interactive)}} \\
    \addlinespace[0.2em]
    
    \textit{Static Hazards \& Geometry} &
    \begin{itemize}[leftmargin=*, nosep, before=\vspace{-0.3\baselineskip}, after=\vspace{-0.3\baselineskip}]
        % Section 3.1: Minimally Invasive Filters (Optimization)
        \item \textbf{Task-Preserving Filter (Minimally Invasive):} \cite{strasser2024collision}$[\mathcal{F}]$   
          \item \textbf{Task-Preserving Filter (Switching):}  \cite{agrawal2024gatekeeper}$[\mathcal{F}]$ 
    \end{itemize} &
    \begin{itemize}[leftmargin=*, nosep, before=\vspace{-0.3\baselineskip}, after=\vspace{-0.3\baselineskip}]
        % Section 4.1: From Min-Max to Contingency MPC
        \item \textbf{Contingency MPC:} \cite{hardy2010contingency, hardy2013contingency, tas2018decision, phiquepal2021control}$[\mathcal{P}]$, \cite{zheng2025safe}$[\mathcal{F}]$ 
        \item \textbf{Robust Min-Max MPC:} \cite{batkovic2021robust}$[\mathcal{F}]$
        \item \textbf{Scenario-Tree MPC:} \cite{Schildbach2015}$[\mathcal{P}]$
    \end{itemize} \\
    
    \addlinespace[0.5em]

    \textit{Interactive Agents} &
    \begin{itemize}[leftmargin=*, nosep, before=\vspace{-0.3\baselineskip}, after=\vspace{-0.3\baselineskip}]
        % Section 3.1: Switching Filters & Shielding
        \item \textbf{Task-Preserving Filter (Switching):} \cite{he2021rule}$[\mathcal{F}]$, \cite{leung2020infusing}$[\mathcal{F}]$
        \item \textbf{Belief-Aware Shield:} \cite{ hu2023deception}$[\mathcal{P}]$
        % Section 3.2: Fail-Safe Supervision
        \item \textbf{Fail-Safe Supervision (MRC):} \cite{Althoff2013, Magdici2016, Pek2021Failsafe,schurmann2017ensuring,Brudigam2023}$[\mathcal{F}]$
        \item \textbf{Fail-Safe Supervision (Shared Control):} \cite{Xue2022Override,Xue2023Shared}$[\mathcal{E}]$
           \item \textbf{Task-Preserving Filter (Shielding):} \cite{wang2024safe, raeesi2025safe}$[\mathcal{F}]$, \cite{oh2025safety}$[\mathcal{E}]$
    \end{itemize} &
    \begin{itemize}[leftmargin=*, nosep, before=\vspace{-0.3\baselineskip}, after=\vspace{-0.3\baselineskip}]
        % Section 4.1: Branch MPC
        \item \textbf{Contingency MPC (Branching):} \cite{schweidel2022driver,chen2023invariant, hu2024active, geurts2025contingency,bouzidi2025reachability,lin2025contingency,yang2025safe}$[\mathcal{F}]$, \cite{zhan2016non,qiu2020latent,bouzidi2024motion,chen2022interactive, li2023marc, wang2023interactive,mustafa2024racp}$[\mathcal{P}]$
        % Section 4.2: Game-Based
        \item \textbf{Game-Based Contingency:} \cite{zhang2024automated}$[\mathcal{F}]$, \cite{ma2025trajectory, huang2026fast, peters2024contingency}$[\mathcal{P}]$
        % Section 4.3: Learning-Based
        \item \textbf{Learning-Based Contingency:} \cite{rhinehart2021contingencies, tolstaya2021identifying,cui2021lookout, chen2023tree,liu2024reasoning,zhong2025coplanner}$[\mathcal{E}]$
    \end{itemize} \\ 
    \addlinespace[0.5em]

    \textit{Occlusion \& Hidden State} &
    \begin{itemize}[leftmargin=*, nosep, before=\vspace{-0.3\baselineskip}, after=\vspace{-0.3\baselineskip}]
        % Section 3.1: Switching (Safe Kernel)
        \item \textbf{Task-Preserving Filter (Switching):} \cite{bajcsy2019efficient}$[\mathcal{F}]$ \cite{chen2021safe}$[\mathcal{F}]$ 
    \end{itemize} &
    \begin{itemize}[leftmargin=*, nosep, before=\vspace{-0.3\baselineskip}, after=\vspace{-0.3\baselineskip}]
        % Section 4.1: Contingency MPC
        \item \textbf{Contingency MPC (Occlusion):} \cite{nyberg2025hope, zheng2025safe, zheng2025occlusion}$[\mathcal{F}]$, \cite{zheng2025oacp}$[\mathcal{P}]$
        % Section 4.2: Game-Based
        \item \textbf{Game-Based Contingency:} \cite{Zhan-RSS-21}$[\mathcal{F}]$; \cite{qiu2024inferring}$[\mathcal{P}]$
        % Section 4.3: Learning-Based
        \item \textbf{Learning-Based (Active):} \cite{packer2023anyone}$[\mathcal{E}]$ 
    \end{itemize} \\
    \addlinespace[0.5em] 

    \textit{Semantic \& OOD Anomalies} &
    \begin{itemize}[leftmargin=*, nosep, before=\vspace{-0.3\baselineskip}, after=\vspace{-0.3\baselineskip}]
        \item \textbf{Task-Preserving Filter (Latent Safety):} \cite{seo2025uncertainty}$[\mathcal{P}]$
    \end{itemize} &
    \begin{itemize}[leftmargin=*, nosep, before=\vspace{-0.3\baselineskip}, after=\vspace{-0.3\baselineskip}]
        \item \textbf{Learning-Based (Foundation Models):} \cite{sinha2024real, ganai2025real}$[\mathcal{P}]$
        \item \textbf{Dynamic Branching (Deferred):} \cite{elango2025deferred}$[\mathcal{P}]$
    \end{itemize} \\

    \midrule
    %=================================================================
    % PART 2: INTERNAL CONTINGENCIES
    %=================================================================
    \multicolumn{3}{@{}l}{\textbf{\color{arblue}II. Internal Contingencies (Ego-System Faults \& Policy Failure)}} \\
    \addlinespace[0.2em]

    \textit{Sensor \& Localization Failure (e.g., GNSS signal loss, LiDAR faults, Camera blindness)} &
    \begin{itemize}[leftmargin=*, nosep, before=\vspace{-0.3\baselineskip}, after=\vspace{-0.3\baselineskip}]
        \item \textbf{Task-Preserving Filter (Minimally Invasive):}    \cite{singletary2022onboard}$[\mathcal{F}]$,
        % Section 3.2: Discrete Logic
        \item \textbf{Task-Preserving Filter (Adaptive):} \cite{yun2025atom}$[\mathcal{P}]$
        % Section 3.1: Switching
        \item \textbf{Task-Preserving Filter (Switching):} \cite{laine2020eyesclosed}$[\mathcal{F}]$
        % Section 3.2: Specific Strategies
        \item \textbf{Fail-Safe Supervision (Fallback):} \cite{Xue2019Virtual, Xue2018Fallback,jonasson2020}$[\mathcal{E}]$, \cite{chakraborty2025system,Viana2022}$[\mathcal{E}]$ \item \textbf{Fail-Safe Supervision (Formal Logic):} \cite{Krook2019, krook2020formal}$[\mathcal{F}]$ 
    \end{itemize} &
    \begin{itemize}[leftmargin=*, nosep, before=\vspace{-0.3\baselineskip}, after=\vspace{-0.3\baselineskip}]
        \item \textit{(Typically handled via reactive approaches rather than pre-planned branches)}
    \end{itemize} \\
    
    \addlinespace[0.5em]

    \textit{Actuation \& Dynamics Faults (e.g., Tire blowouts, Electric motor malfunctions)} &
    \begin{itemize}[leftmargin=*, nosep, before=\vspace{-0.3\baselineskip}, after=\vspace{-0.3\baselineskip}]
        % Section 3.1: Minimally Invasive (Fault-Tolerant) 
        \item \textbf{Task-Preserving Filter (Fault-Tolerant):} \cite{zhang2025safe}$[\mathcal{F}]$ 
        \item \textbf{Task-Preserving Filter (Model Reliability Monitor):} \cite{fisac2019general}$[\mathcal{F}]$
        % Section 3.2: System Level
        \item \textbf{Fail-Safe Supervision (Control Alloc.):} \cite{yue2019automated,yu2019fallback, boudali2018emergency, khelladi2020emergency, duerr2020realtime,lodder2023optimization}$[\mathcal{E}]$, \cite{lee2022adaptive}$[\mathcal{F}]$
       \item \textbf{Shared Control (Faults):} \cite{li2023novel,li2020shared}$[\mathcal{E}]$ % 
        \item \textbf{Fail-Safe Supervision (System Redundancy):} \cite{julitz2023computer}$[\mathcal{P}]$; \cite{mercedesbenz2022,pechinger2020hardware}$[\mathcal{E}]$
    \end{itemize} &
    \begin{itemize}[leftmargin=*, nosep, before=\vspace{-0.3\baselineskip}, after=\vspace{-0.3\baselineskip}]
        % Section 4.1: Contingency MPC (State-Based)
        \item \textbf{Contingency MPC (Friction):} \cite{alsterda2019contingency, alsterda2021contingency}$[\mathcal{F}]$
    \end{itemize} \\

    \bottomrule
\end{tabularx}
  
     % \vspace{-0pt}
\end{table*}

\subsection{{Closed-Loop Mode Transitions}}
\label{subsec:closed_loop_mode_transition} 
{
Scenario-tree-based Proactive contingency planning has an
information-feedback structure because future contingency actions are
conditioned on information revealed during execution rather than executed
unconditionally. Within the framework of
Section~\ref{sec:problem_formulation}, the logical mode evolves
according to the transition map $T_k$, whereas the observations summarized
by the information set $\mathcal I_k$ determine when that mode can be
resolved and which mode-conditioned continuation is selected.
For example, in the left-turn scenario considered
in~\cite{alsterda2019contingency}, the planner reasons about alternative
future conditions, such as normal and low-friction road surfaces, and
proactively prepares corresponding contingency plans. The planner then
identifies the relevant mode from online observations and selects the
corresponding mode-conditioned continuation.
}

{
Beyond execution uncertainty, contingency planning becomes
interaction-aware when the behaviors and responses of other agents are
explicitly incorporated into contingency generation and selection.
For example, in a lane-merging task, an AV may maintain
contingency plans corresponding to different traffic evolutions, such as
a neighboring vehicle yielding or accelerating. The planner selects the
corresponding continuation based on observed traffic behavior.
Furthermore, we can categorize interaction-aware contingency planning into
passive and active approaches, depending on whether the behaviors of other
agents are treated as external predictions or modeled as action-dependent
responses within the planning process. This distinction refines rather
than replaces the Reactive--Proactive taxonomy: Reactive and Proactive
Safety are distinguished by whether the relevant contingency mode is
resolved, whereas passive and active approaches differ in whether the ego
vehicle's actions influence the evolution or resolution of unresolved
interaction modes.
}

{
In passive contingency planning, the planner relies on predicted agent
modes, trajectories, or intent distributions to construct contingency
branches and optimize the future actions of the ego vehicle. Although
such approaches account for multiple possible interaction outcomes, the
influence of the ego vehicle's decisions on the behaviors of other
agents is not explicitly modeled. Within
Eq.~\eqref{eq:mode_logic}, this corresponds to representing the
influence of other agents on the mode transition entirely through the
external factor $v_k$. Consequently, the scenario tree is typically
constructed using externally estimated probabilities of other agents'
future behaviors.
}

{
In contrast, active contingency planning~\citep{hu2024active} models
future agent responses as dependent on the ego vehicle's decisions. Within the framework of Section~\ref{sec:problem_formulation}, $v_k$
remains an external factor outside the ego vehicle's control, while part
of the response of another agent may be represented through the
dependence of $T_k(x_k,\sigma_k,v_k)$ on the ego state $x_k$. For
example, whether another vehicle cuts in may depend on the ego vehicle's
position and speed, in addition to driver-specific factors represented
by $v_k$. This dependence may affect both the
construction of the scenario tree and, in probabilistic formulations, its
branch weights. Since the ego control affects the future state trajectory, the planner
may thereby influence the likelihood of subsequent interaction outcomes. A related
class of methods is game-theoretic contingency planning~\citep{peters2024contingency},
which models mutual interaction as a strategic decision problem in which
each agent accounts for the possible responses of others.
}

{
This interaction model is reflected in both the logical-mode evolution
and the information structure of the scenario tree. While several
interaction outcomes remain compatible with the available information,
branches sharing the same information history are coupled by the
non-anticipativity constraint in~\eqref{eq:rcbdp_nonant}. Different
realizations of $v_k$ may induce distinct mode transitions through $T_k$,
whereas the evolving information set $\mathcal I_k$ determines when the
corresponding branches become distinguishable. Specifically,
$\tau(b,b')$ denotes the first time at which branches $b$ and $b'$ can be
distinguished; only then may their mode-conditioned controls diverge. In
active formulations, the ego vehicle's actions may also affect both the
likelihood of future outcomes and the time at which they become
distinguishable. For a single split, this time is represented by the
branching time $t_b$, whose determination is discussed in
Section~\ref{subsec:dynamic_branching}. Such interaction modeling enriches
Proactive planning but does not by itself establish safety, which also
depends on model validity, scenario coverage, and suitable terminal or
runtime safety conditions.
}

% {At the branching point, different event realizations induce different discrete mode transitions through Eq.~\eqref{eq:mode_logic}. 
% For example, an event indicating that the surrounding vehicle sufficiently decelerates may transition the planner from the unresolved mode \(\sigma^U\) to a conflict-clear mode, whereas an event indicating that the surrounding vehicle enters or threatens the conflict region may transition it to a conflict-blocked mode. 
% The shared trunk therefore represents a period of unresolved interaction, while the branches represent resolved logical modes with different safety specifications. 
% How to determine this branching point is discussed next in Section~\ref{subsec:dynamic_branching}.}

\subsection{Dynamic Branching Point Determination}
\label{subsec:dynamic_branching}
Determining when to branch is a central challenge in Proactive contingency planning, as it directly balances safety assurance against operational efficiency.  
Early branching can reduce conservatism, but introduces significant safety risks if critical uncertainty remains unresolved by the branching time. In contrast, branching late enhances robustness but typically results in excessive conservatism, as the vehicle must maintain a valid fallback for conflicting futures over an extended horizon.  { Theoretically, the branching time $t_b$ provides a useful connection between Reactive and Proactive formulations. When the relevant mode is
already resolved, the scenario tree collapses to the corresponding mode-conditioned branch; equivalently, setting $t_b=0$ removes the shared trunk and recovers the information structure underlying the Reactive formulation. Conversely, postponing branching beyond the planning horizon requires all modeled mode alternatives to share a common control sequence throughout the horizon, yielding an open-loop robust formulation over the modeled modes}. To mediate this trade-off, recent work has shifted toward adaptive strategies that dynamically determine the branching time based on real-time scene context and the level of prediction uncertainty.

Heuristic-based strategies offer computationally efficient solutions by triggering branches through interpretable indicators. One common class monitors spatial divergence across predicted trajectories. For instance, the MARC framework \citep{li2023marc} identifies the latest time at which the ego vehicle's trajectories for different behavior modes remain within a predefined deviation threshold, defining a dynamic branching point from this "scene-level divergence." Another class leverages information-theoretic criteria to postpone decisions until predicted futures become statistically distinguishable. This concept was demonstrated by \cite{tas2018decision} and later refined by \cite{bouzidi2024motion}, which used the Bhattacharyya distance to quantify the separation between scenario probability distributions. In the Bhattacharyya-based approach, the vehicle continues on a common path until the predicted outcome distributions diverge beyond a threshold, triggering a branch only once the futures are sufficiently distinct. While these heuristic triggers are lightweight and interpretable, they depend on carefully tuned threshold values, which may limit their robustness and generalizability across scenarios.

A more formal approach defines the branching point based on safety boundaries, focusing on the latest possible moment at which a decision can be made without compromising future safety. This is typically achieved through reachability analysis, which computes the maximum decision latency, the last moment at which the shared trunk trajectory still allows a safe and feasible continuation for all possible future scenarios \citep{bouzidi2025reachability}.  
% Theoretically, this reachability-based boundary corresponds to the edge of the robust maximal control-invariant set (accounting for all logical modes), defining the state space region where the system retains valid recourse against the worst-case realization of uncertainty. 
While providing strong safety guarantees under the corresponding modeling assumptions, this approach can be conservative as it is based on worst-case feasibility.
To operate more efficiently within this safe window, \cite{elango2025deferred} cast the branching point itself as a decision variable within trajectory optimization. A key contribution in this direction is Deferred-Decision Trajectory Optimization, which jointly optimizes both the continuous trajectory and the discrete branching time. This allows the planner to "wait and see", dynamically committing to a branch only when informative cues appear. By integrating uncertainty-aware cost terms that implicitly model the value of information, these formulations quantify the benefit of delayed decision-making while maintaining robustness.

Despite these advances, key challenges remain. The computational complexity of optimizing over dynamically branching scenario trees hinders real-time deployment, especially in dense, multi-agent settings. Furthermore, these methods rely heavily on the accuracy and calibrated uncertainty of upstream prediction modules. Promising future directions lie in hybrid frameworks that use formal reachability analysis to define a certified safe decision window, within which advanced optimization or learning-based planners can flexibly determine the optimal time to branch. This hybrid approach enables a principled trade-off between safety, efficiency, and scalability in dynamic environments.

\section{{Summary and Future Prospects}}\label{sec:future} 
In this paper, we presented contingency planning as a cornerstone for the safe and reliable deployment of high-autonomy, safety-critical AVs. This capability is essential for managing multimodal uncertainty while maintaining operational safety. We introduced a logic-conditioned problem formulation and a comprehensive taxonomy of contingency planning approaches, categorizing existing methods into {Reactive Safety} and {Proactive Safety} paradigms. For each class, we discussed foundational concepts, representative algorithms, safety guarantees, and deployment challenges. 
Table~\ref{tab:contingency_comparison} provides a comprehensive classification of the literature reviewed in this survey. Complementing this taxonomy, Table~\ref{tab:comparison} synthesizes the trade-offs between paradigms regarding safety guarantees,  behavioral conservatism, computational complexity, scalability, and adaptivity. 

Across these methodologies, a common goal emerges: enabling safe decision-making under uncertainty—the ability of AVs to respond robustly to rare and unpredictable situations while preserving natural and efficient driving behavior. Strategies vary significantly: some emphasize pre-validated fallback trajectories as safety filters, while others adopt scenario branching and dynamic replanning. Recent methods increasingly leverage learning-based prediction and planning to enhance behavioral richness. Despite their diversity, the field is converging toward hybrid solutions that combine multiple paradigms to overcome individual limitations.

However, bridging the gap between formal safety guarantees and real-world robustness remains a core challenge. We conclude by identifying critical areas that currently limit the practical deployment of contingency planning methods.

\subsection{Robust and Scalable Modeling}
A critical challenge lies in building models that are both robust and scalable across diverse driving scenarios. Many existing planners depend on simplified dynamics, such as kinematic bicycle or point-mass models, which typically overlook important physical phenomena like roll, pitch, terrain interaction, or actuator degradation. {These low-fidelity approximations reduce robustness, especially in unstructured environments or under fault conditions where accurate modeling is crucial for safe fallback behavior.}

To ensure formal safety guarantees, many methods assume strict structural priors. Specifically, {many commonly used CBF-QP implementations typically assume control-affine
dynamics so that the barrier condition is affine in the control input.
HJ reachability typically requires regularity assumptions, such as
Lipschitz-continuous dynamics, to ensure well-posedness and support
numerical approximation. Additionally, learning-based certificates typically assume representative offline datasets, consisting of independent and identically distributed samples, together with bounded uncertainty, to derive statistical guarantees. These assumptions typically fail in complex, high-dimensional scenarios, such as deformable terrain or adversarial traffic.} A promising direction involves detecting modeling errors and combining physics-informed priors with real-time data-driven adaptation. Emerging techniques, such as online system identification, neural differential equations, and latent state abstractions, enable planners to adapt to environmental variability while retaining physical interpretability. Although higher-fidelity models can be computationally demanding, methods like model-order reduction and accelerated simulation help mitigate this cost.

{System-level modeling must be complemented by explicit treatment of uncertainty from upstream modules, particularly perception and
prediction.} Many planners assume reliable maps, accurate intent estimation, and calibrated forecasting of other agents. In open-world deployments, these white-box assumptions are typically violated. While black-box methods (e.g., scenario sampling) attempt to circumvent these modeling errors by relying on data-driven distributions, they trade the "validity gap" for a "coverage gap," suffering from sample inefficiency and the inability to guarantee safety against rare, unobserved events~\citep{corso2021survey}. 

To address this gap, {researchers are increasingly adopting learned world models and calibrated uncertainty estimates while actively monitoring assumption violations. Effective strategies include: 
(i) detecting distribution shifts and adjusting branching strategies accordingly~\citep{sinha2024real,ganai2025real};
(ii) embedding intent uncertainty diagnostics to guide belief updates and information-seeking actions~\citep{mustafa2024racp,liu2024reasoning}; and
(iii) utilizing confidence-based triggers to activate conservative fallback policies~\citep{seo2025uncertainty} and conformal prediction to provide 
finite-sample coverage under the corresponding calibration
assumptions~\citep{lindemann2025formal}. 
These mechanisms reduce the brittleness of high-level assumptions without defaulting to overly pessimistic, worst-case designs.}

\subsection{Synthesis of  Invariant Sets for Scalable Learning-Based Safety}  
Control-invariant sets define regions of the state space where safety-preserving controls are guaranteed to exist. They form the foundation for fallback policy design and  {are fundamental for establishing forward safety and recursive feasibility in learning-based systems
under the corresponding invariance assumptions}. In this context, invariant sets are not auxiliary tools but fundamental building blocks for scalable and verifiable safety synthesis in high-dimensional, partially known, or data-driven environments. {
For interconnected systems, compositional construction of
input-to-state safe (ISSf) barrier functions from subsystem
certificates under cyclic small-gain conditions provides a
related perspective on scalable safety certification
~\citep{wang2025construction}.
}

A major challenge is synthesizing such sets under relaxed assumptions, where complete model knowledge, low dimensionality, or tractable dynamics can no longer be assumed. This challenge is especially significant for high-dimensional AVs under uncertain terrains or multi-agent human-robot interactions. Classical tools like reachability analysis or convex optimization typically face scalability limitations in these contexts.

 Recent work investigates implicit safety representations, such as value functions or neural approximators, which offer greater representational flexibility but require additional
verification to establish formal guarantees \citep{seo2025uncertainty,bansal2021deepreach}. In particular, latent value representations, learned through neural ODEs, encoder–decoder architectures, or contrastive objectives, offer compact abstractions that support policy generalization in complex, partially observable domains. 

Complementing these value-based approaches, conditional generative modeling frameworks (e.g., diffusion models, flow-matching networks) offer a novel paradigm for implicitly characterizing high-dimensional safe manifolds through sampling. Rather than constructing geometric boundaries, these methods integrate CBFs or safety constraints directly into the generative vector fields or denoising processes \citep{dai2025safe, huang2025sad, xiao2025safediffuser}. {By aligning the generative flow with invariance conditions, this approach transforms the safety synthesis problem from geometric computation to constrained distribution learning. This produces a generative prior for sampling constraint-aware trajectories in high-dimensional spaces. However, formal invariance and recursive feasibility still require separate verification.}
 
\subsection{Runtime Safety Assurance: From Abstraction to Execution}  
Achieving scalable and generalizable safety guarantees remains one of the central challenges in contingency planning. While the synthesis methods discussed above provide candidate safety envelopes (e.g., neural value functions or generative manifolds), deploying these abstractions under real-world uncertainty introduces a distinct verification gap. Although risk-constrained formulations can impose probabilistic safety constraints, extending safety assurance to long-horizon multimodal tasks under perception uncertainty is nontrivial. 

The core difficulty lies in translating formal safety abstractions into runtime mechanisms that remain effective under perception noise, sensor latency, and uncertain human interactions. AVs operate in highly dynamic environments where failures in perception or actuators can be difficult to detect and mitigate. Therefore, runtime monitors should detect faults and trigger intervention before safety margins are exhausted, accounting for sensing, computation, and actuation delays.  Even when failures are detected, planners may not have sufficient time to replan a safe trajectory under degraded information conditions~\citep{chakraborty2025system}. 

One promising direction is to use context-conditioned control-invariant sets that adapt to latent environmental factors~\citep{seo2025uncertainty, oh2025safety} for both runtime monitoring and fallback planning. For runtime monitoring, learned implicit representations (e.g., value functions $V(x)$) can track proximity to constraint boundaries and trigger corrective actions when violations are imminent. As planning primitives, they define verifiable terminal regions that support adaptive, certifiable contingency trajectories. Integrating these dual roles within hybrid architectures, linking design-time synthesis, runtime monitoring, and fallback control, builds a continuous pipeline from modeling to execution, forming the basis for verifiable safety assurance. 

\subsection{ Socially-Aware and Ethical Contingency Planning}
Beyond physical safety, contingency planning must also account for the social and ethical dimensions of autonomous driving. AVs operate in highly interactive environments, where decisions depend on interpreting human intent, complying with traffic standards, and maintaining mutual predictability~\citep{krugel2024risk}. Effective planners must reason about behavioral context to ensure that actions are interpretable and acceptable to surrounding road users.

Embedding ethical reasoning introduces new challenges. Real-world driving presents moral trade-offs, such as balancing risk among participants or minimizing unavoidable harm, that require formalization into tractable specifications~\citep{evans2020ethical}. Recent research has explored symbolic logic constraints, socially compliant utility functions, and multi-objective decision frameworks to encode principles of fairness, courtesy, and harm minimization. {Fundamentally, these considerations reshape how safety specifications are defined, transforming them from rigid hard constraints into prioritized objectives. Unlike the fixed state constraints assumed in the standard hybrid framework reviewed in this survey, ethical reasoning often necessitates solving hierarchical optimization problems over conflicting specifications (e.g., prioritizing preserving life over adhering to traffic rules). Consequently, while the core Reactive and Proactive paradigms remain applicable as the foundational execution layer, ethical contingency planners should be viewed as advanced decision modules built on top of these methodologies, dynamically modulating constraints to satisfy complex social contracts.} 

Ultimately, ethically aligned contingency planning aims to connect formal safety guarantees with societal expectations.
Realizing this goal requires not only reliable and adaptive planning but also transparency and fairness across cultural and situational contexts. This represents a key frontier in building socially compatible and trustworthy autonomous systems.

\subsection{{Standardized Benchmarking and Evaluation}}
\label{subsec:benchmarking}
To transition contingency planning from theoretical formulations to deployable systems, the community must adopt standardized benchmarking that moves beyond ad hoc simulations. Critically, the choice of evaluation platform must be dictated by the nature of the contingency.

\textbf{Platform Selection by Contingency Type}
For contingencies involving fundamental kinematic limits and rule compliance, benchmarks must  {provide explicit analytical models (e.g., white-box dynamics) to rigorously validate constraint satisfaction}. Platforms such as \textbf{CommonRoad} \citep{althoff2017commonroad} provide standardized, composable scenarios ideal for assessing the feasibility and conservatism of Reactive filters without the confounding variables of perception noise. Conversely, contingencies stemming from intent uncertainty require data-driven evaluation. Log-replay platforms like \textbf{Waymax} \citep{gulino2023waymax} and \textbf{nuPlan} \citep{caesar2021nuplan} offer large-scale, real-world trajectory data, which is essential for validating whether Proactive branching strategies can robustly handle realistic human unpredictability. Finally, for "zero-order" observability failures or vehicle dynamics faults, high-fidelity physics engines are necessary. \textbf{CARLA} \citep{dosovitskiy2017carla}, supported by Unreal Engine physics, allows for the injection of raw sensor noise and mechanical degradation, enabling the evaluation of empirical fail-safe triggers against photorealistic distribution shifts.

\textbf{{A Hierarchy of Standardized Metrics}}
Standard safety metrics, such as collision rate, are insufficient for evaluating contingency planners, as catastrophic failures are rare by definition. To rigorously dissect performance, evaluation must prioritize {worst-case safety margins}. Rather than reporting abstract value functions $V(x)$ that vary across methods, it is ideal for benchmarks to additionally report {the minimum physical signed distance to the failure set boundary during critical events.}

% This provides a universal measure of robustness for invariant sets. 

Complementing this, the cost of conservatism can be assessed by measuring the performance degradation (e.g., velocity loss) relative to a non-contingency oracle in nominal scenarios. For Proactive planners, {we recommend reporting the branch realization ratio—the proportion of planned contingency maneuvers that are actually executed.} This metric helps distinguish between prudent anticipation and excessive caution, avoiding the pitfall of penalizing planners for hedging against risks that do not materialize. Conversely, the intrusiveness of Reactive filters is often captured {via intervention frequency and induced jerk, ensuring that safety enforcement does not destabilize the vehicle or degrade passenger comfort.}  Finally, {for fail-safe supervision, a critical metric involves the time to MRC, quantifying the system's latency in transitioning from a fault injection to a verified safe state. }
 
 Synthesizing these disparate indicators, ranging from formal safety bounds to empirical comfort measures,  into a unified evaluation framework remains a significant open challenge. Simply aggregating them into a scalar score risks obscuring critical nuances. A more rigorous path forward {may lie in characterizing the Pareto frontier between safety (e.g., minimum distance to failure) and operational efficiency (e.g., branch realization).} Ultimately, establishing this multifaceted metric suite serves as a necessary precursor to rigorous cross-paradigm comparison and certification.

\section*{Declaration of Generative AI and AI-assisted technologies in the manuscript preparation process}

During the preparation of this work, the author(s) used Google's Gemini and OpenAI's ChatGPT in order to improve the language and readability of the manuscript. After using this tool/service, the author(s) reviewed and edited the content as needed and take(s) full responsibility for the content of the published article.

\bibliographystyle{cas-model2-names}
% \begin{ack}
% Place acknowledgments here.
% \end{ack}
% \small
\bibliography{egbib}             % bib file to produce the bibliography
                                                     % with bibtex (preferred)
                                                   
%\begin{thebibliography}{xx}  % you can also add the bibliography by hand

%\bibitem[Able(1956)]{Abl:56}
%B.C. Able.
%\newblock Nucleic acid content of microscope.
%\newblock \emph{Nature}, 135:\penalty0 7--9, 1956.

%\bibitem[Able et~al.(1954)Able, Tagg, and Rush]{AbTaRu:54}
%B.C. Able, R.A. Tagg, and M.~Rush.
%\newblock Enzyme-catalyzed cellular transanimations.
%\newblock In A.F. Round, editor, \emph{Advances in Enzymology}, volume~2, pages
%  125--247. Academic Press, New York, 3rd edition, 1954.

%\bibitem[Keohane(1958)]{Keo:58}
%R.~Keohane.
%\newblock \emph{Power and Interdependence: World Politics in Transitions}.
%\newblock Little, Brown \& Co., Boston, 1958.

%\bibitem[Powers(1985)]{Pow:85}
%T.~Powers.
%\newblock Is there a way out?
%\newblock \emph{Harpers}, pages 35--47, June 1985.

%\bibitem[Soukhanov(1992)]{Heritage:92}
%A.~H. Soukhanov, editor.
%\newblock \emph{{The American Heritage. Dictionary of the American Language}}.
%\newblock Houghton Mifflin Company, 1992.

%\end{thebibliography}

% \appendix
% \section{A summary of Latin grammar}    % Each appendix must have a short title.
% \section{Some Latin vocabulary}              % Sections and subsections are supported  
%                                                                          % in the appendices.
\end{document}